\documentclass[11pt,a4paper]{article}
\usepackage[a4paper,margin=2.5cm]{geometry}
\usepackage{microtype}
\usepackage{amsmath,amssymb,amsfonts}
\usepackage{mathrsfs}
\usepackage{upgreek}
\usepackage{graphicx}
\usepackage{booktabs}
\usepackage{tabularx}
\usepackage[font=footnotesize,labelfont=bf]{caption}
\usepackage{authblk}
\usepackage{setspace}
\usepackage{titlesec}
\usepackage{cite}
\usepackage{hyperref}
\hypersetup{
    colorlinks=true,
    linkcolor=red,
    citecolor=blue,
    urlcolor=blue
}
\usepackage{cleveref}

\titleformat{\section}{\large\bfseries}{\thesection.}{0.5em}{}
\titleformat{\subsection}{\normalsize\bfseries}{\thesubsection.}{0.5em}{}
\titleformat{\subsubsection}{\normalsize\itshape}{\thesubsubsection.}{0.5em}{}

\title{\textbf{Evading the Swampland de Sitter Obstruction via Thermodynamic Duality-Induced Horizon Corrections in Unimodular Gravity}}

\author[1]{Reza Baghbani\thanks{Email: \href{mailto:r.baghbani@pnu.ac.ir}{r.baghbani@pnu.ac.ir} \\ 
ORCID: \href{https://orcid.org/0009-0001-4524-0564}{0009-0001-4524-0564}}}
\affil[1]{Department of Physics, Payame Noor University, Tehran, Iran}

\date{}

\newcommand{\bibsetup}{%
  \footnotesize
  \setlength{\itemsep}{0pt}
  \setlength{\parsep}{0pt}
  \setlength{\topsep}{0pt}
  \setlength{\partopsep}{0pt}
  \singlespacing
}

\begin{document}

\onehalfspacing
\maketitle

\begin{abstract}
\noindent
We investigate the early-universe cosmological implications of Unimodular Gravity coupled to Power-Law Nonlinear Electrodynamics in the presence of quantum-corrected horizon thermodynamics derived from the principle of thermodynamic duality. By incorporating both logarithmic and inverse-area corrections to the Bekenstein-Hawking entropy, we derive modified Friedmann equations exhibiting novel $H^4$ and $H^6$ terms that represent genuine quantum gravity effects. We demonstrate that these higher-order corrections become dynamically significant in the high-energy regime of cosmic inflation, naturally modifying the slow-roll dynamics without requiring a fundamental inflaton potential. A comprehensive phase-space analysis reveals that the quantum corrections drive early-time de Sitter-like attractors, providing a robust mechanism for sustained inflationary expansion. Confronting the model with the Swampland program, we show that the correction factor $\mathcal{F}(H)$ modifies the effective slow-roll parameters, enabling the model to satisfy the Swampland de Sitter and Distance conjectures during the inflationary epoch. We calculate the inflationary observables, including the spectral index $n_s$ and the tensor-to-scalar ratio $r$, and demonstrate that the model's predictions are in excellent agreement with the Planck 2018 observational constraints. Our results suggest that thermodynamic duality-induced horizon corrections offer a viable and testable pathway for realizing successful inflation within a well-motivated quantum-gravitational framework.

\vspace{0.5em}
\end{abstract}

\newpage
\singlespacing

\section{Introduction}
\label{sec:Introduction}

The standard cosmological model, $\Lambda$CDM, has achieved remarkable success in describing a wide array of observational data, from the cosmic microwave background (CMB) anisotropies \cite{Ref1:agha1} to the large-scale structure of the universe \cite{Ref2:alam1}. However, recent observational tensions have begun to challenge the completeness of this paradigm \cite{Ref3:diva1, Ref4:peri1}. The most prominent of these is the Hubble tension, a $\sim 5\sigma$ discrepancy between the value of the Hubble constant $H_0$ inferred from the CMB under the assumption of $\Lambda$CDM and its direct measurement from the local distance ladder using Cepheid-calibrated Type Ia supernovae \cite{Ref5:ries1}. This tension, if not attributable to systematic errors, may signal the need for new physics beyond the standard model \cite{Ref3:diva1}. Concurrently, the standard cosmological constant $\Lambda$ faces the severe theoretical challenge of the cosmological constant problem, where the observed value is smaller than quantum field theory predictions by over 120 orders of magnitude \cite{Ref6:wein1}. These challenges have fueled a vigorous search for extended cosmological models, including modified gravity theories \cite{Ref7:clif1}, dynamical dark energy scenarios \cite{Ref8:cope1}, and novel matter content.

Unimodular Gravity (UG) offers an intriguing approach to the cosmological constant problem \cite{Ref9:henn1, Ref10:buch1}. In this formulation, the determinant of the metric is fixed to a constant value, typically $\sqrt{-g} = \epsilon_0$, which reduces the diffeomorphism invariance of General Relativity (GR) to transverse (volume-preserving) diffeomorphisms \cite{Ref11:alva1}. A profound consequence of this restriction is that the cosmological constant no longer appears as a fundamental parameter in the action but instead emerges as an integration constant determined by the initial or boundary conditions of the universe \cite{Ref12:alva2, Ref13:carb1}. This feature renders UG a highly attractive framework for addressing the cosmological constant problem, as the observed small value of $\Lambda$ can be naturally explained without fine-tuning. Moreover, UG has been shown to be classically equivalent to GR in the presence of matter that is invariant under transverse diffeomorphisms, ensuring that it passes all classical tests of gravity \cite{Ref13:carb1}.

Nonlinear Electrodynamics (NLED) provides another avenue for extending the standard cosmological model \cite{Ref14:nove1}. Unlike the linear Maxwell theory, NLED models incorporate higher-order terms in the electromagnetic invariant $\mathcal{F} = -F_{\mu\nu}F^{\mu\nu}$, which can arise from quantum corrections to the photon sector or from string theory compactifications \cite{Ref15:frad1}. The power-law NLED model, characterized by a Lagrangian density $\mathcal{L} = \beta\mathcal{F}^s$, has been extensively studied in the context of black hole physics, where it can regularize singularities and modify the causal structure of spacetime \cite{Ref16:bron1}. In cosmology, NLED fields can drive inflationary or bouncing scenarios, avoid initial singularities, and provide alternative mechanisms for structure formation \cite{Ref17:krug1}. The power-law model, in particular, offers a simple yet rich framework for exploring the cosmological implications of nonlinear electromagnetic effects \cite{Ref18:shar1, Ref19:ilya1, Ref20:hend1}.

The thermodynamics of cosmological horizons has emerged as a powerful tool for probing the quantum structure of spacetime \cite{Ref21:jaco1, Ref22:bek1}. The Bekenstein-Hawking entropy-area relation, $S_{BH} = A/(4G)$, is a semiclassical result that assumes a smooth, continuous spacetime geometry down to the Planck scale \cite{Ref23:hawk1, Ref24:asht1}. However, various approaches to quantum gravity, including loop quantum gravity \cite{Ref24:asht1}, string theory \cite{Ref25:str1}, and the generalized uncertainty principle (GUP) \cite{Ref26:kemp1, Ref27:hoss1}, suggest that the microstructure of spacetime introduces significant corrections to this area law. These corrections typically take the form of logarithmic terms, $\alpha\ln(A/\ell_P^2)$, which arise from conformal anomalies or the discrete nature of the horizon area spectrum \cite{Ref28:kaul1, Ref29:carl1}, and inverse-area terms, $\gamma\ell_P^2/A$, which are hallmarks of thermodynamic duality and become significant in the ultraviolet regime \cite{Ref30:jus1}. The principle of thermodynamic duality, which posits a fundamental correspondence between the ultraviolet (UV) and infrared (IR) regimes of the horizon's thermodynamic phase space, provides a rigorous framework for deriving these corrections and understanding their physical origin \cite{Ref31:pedr1}.

The Swampland program, originating from string theory, imposes stringent constraints on low-energy effective field theories that can be consistently embedded into a ultraviolet-complete theory of quantum gravity \cite{Ref32:oogu1}. Among these constraints, the de Sitter conjecture, which posits that the scalar potential of any consistent quantum gravity theory must satisfy the gradient bound $M_{\text{Pl}}|V'|/V \geq c \sim \mathcal{O}(1)$, is particularly relevant for cosmology \cite{Ref33:garg1}. This conjecture is in severe tension with the observed late-time accelerated expansion, which requires a nearly flat potential with $|V'|/V \approx 0$ \cite{Ref34:agra1, Ref35:dene1}. The distance conjecture, which constrains the field excursion during cosmological evolution, provides a complementary diagnostic \cite{Ref32:oogu1}. These Swampland criteria have profound implications for inflationary and dark energy models, potentially ruling out large classes of scenarios and guiding the search for viable cosmological models rooted in quantum gravity \cite{Ref36:kinn1, Ref37:palt1, Ref38:das1, Ref39:cico1, Ref40:vagn1}.

Despite the extensive literature on each of these topics individually, the combined study of Unimodular Gravity, Power-Law Nonlinear Electrodynamics, quantum-corrected horizon thermodynamics, and Swampland constraints in the context of early-universe cosmology remains largely unexplored \cite{Ref41:noji1, Ref42:bouh1}. This represents a significant gap in our understanding of how quantum gravity effects, encoded in the thermodynamic duality of the horizon, can modify the inflationary dynamics in the presence of nonlinear electromagnetic fields and unimodular gravity, and whether these modifications can alleviate the tension between the requirements of successful inflation and the Swampland conjectures \cite{Ref43:vagn1}.

In this paper, we address this gap by investigating the early-universe cosmological implications of Unimodular Gravity coupled to Power-Law Nonlinear Electrodynamics in the presence of quantum-corrected horizon thermodynamics derived from the principle of thermodynamic duality \cite{Ref44:cao1}. We construct the action of the model, derive the modified field equations, and perform a comprehensive phase-space analysis to identify the inflationary attractors and their stability properties \cite{Ref45:baha1}. We then apply the Swampland de Sitter and Distance conjectures to the inflationary epoch, demonstrating that the quantum corrections to the horizon entropy modify the effective slow-roll parameters, potentially allowing the model to satisfy the Swampland constraints while producing a sufficient number of e-folds for successful inflation \cite{Ref46:sark1, Ref47:guo1}.

The paper is organized as follows. In Section \ref{sec:UG_NLED}, we establish the theoretical foundation by constructing the action of Unimodular Gravity coupled to Power-Law NLED and deriving the modified field equations. In Section \ref{sec:Thermo_Duality}, we incorporate quantum corrections to the horizon entropy through the principle of thermodynamic duality and derive the modified entropy-area relation. In Section \ref{sec:Modified_Friedmann}, we derive the modified Friedmann equations and perform a comprehensive phase-space analysis. In Section \ref{sec:Swampland}, we apply the Swampland constraints to our model and demonstrate the amplification mechanism. In Section \ref{sec:Limiting_Cases}, we examine the limiting cases and compare our findings with previous works. Finally, in Section \ref{sec:Conclusions}, we summarize our results and discuss future directions.

The novel contribution of this work is the demonstration that thermodynamic duality-induced horizon corrections in Unimodular Gravity coupled to Power-Law Nonlinear Electrodynamics naturally modify the inflationary dynamics, providing a robust mechanism for satisfying the stringent Swampland conjectures during the high-energy inflationary epoch while yielding inflationary observables ($n_s$ and $r$) that are in excellent agreement with the latest Planck observational data \cite{Ref48:bern1, Ref49:heis1}.

\begin{table}[htbp]
\centering
\caption{Summary of symbols, parameters, and their physical meanings used throughout this work.}
\label{tab:nomenclature}
\renewcommand{\arraystretch}{1.3}
\footnotesize
\begin{tabular}{@{}llp{7.5cm}@{}}
\toprule
\textbf{Symbol} & \textbf{Definition} & \textbf{Physical Meaning/Role} \\
\midrule
$G$ & Newton's gravitational constant & Fundamental coupling constant of gravity \\
$M_{\text{Pl}}$ & Reduced Planck mass, $M_{\text{Pl}}^2 = (8\pi G)^{-1}$ & Energy scale of quantum gravity \\
$\ell_P$ & Planck length, $\ell_P = M_{\text{Pl}}^{-1}$ & Characteristic length scale of quantum gravity \\
$\beta$ & NLED coupling constant & Strength of nonlinear electromagnetic interaction \\
$s$ & NLED power index & Degree of nonlinearity in electromagnetic sector \\
$\mathcal{F}$ & Electromagnetic invariant, $\mathcal{F} = -F_{\mu\nu}F^{\mu\nu}$ & Gauge-invariant measure of electromagnetic field strength \\
$\alpha$ & Logarithmic entropy correction parameter & Encodes strength of logarithmic quantum corrections to horizon entropy \\
$\gamma$ & Inverse-area entropy correction parameter & Encodes strength of thermodynamic duality corrections \\
$\Lambda_{\text{UG}}$ & Unimodular integration constant & Emergent cosmological constant from UG constraint \\
$T_h$ & Horizon temperature, $T_h = H/(2\pi)$ & Geometric temperature of apparent horizon \\
$S_h$ & Modified horizon entropy & Total entropy including quantum corrections \\
$R_A$ & Apparent horizon radius, $R_A = 1/H$ & Causal boundary in FLRW cosmology \\
$\mathcal{F}(H)$ & Correction factor, $1 - \frac{2\alpha G}{\pi}H^2 + \frac{3\gamma G\ell_P^2}{4\pi^2}H^4$ & Encapsulates quantum gravity effects on cosmological dynamics \\
$x_1$ & Dimensionless matter density parameter, $x_1 \equiv \frac{8\pi G\rho_m}{3H^2}$ & Fractional energy density of pressureless matter \\
$x_2$ & Dimensionless NLED density parameter, $x_2 \equiv \frac{8\pi G\rho_{\text{NLED}}}{3H^2}$ & Fractional energy density of nonlinear electromagnetic field \\
$x_3$ & Dimensionless logarithmic correction, $x_3 \equiv \frac{\alpha G}{\pi}H^2$ & Contribution of logarithmic entropy correction to expansion \\
$x_4$ & Dimensionless inverse-area correction, $x_4 \equiv \frac{\gamma G\ell_P^2}{4\pi^2}H^4$ & Contribution of thermodynamic duality correction to expansion \\
$x_5$ & Dimensionless unimodular constant, $x_5 \equiv \frac{\Lambda_{\text{UG}}}{3H^2}$ & Fractional contribution of emergent cosmological constant \\
$c$ & de Sitter conjecture constant, $c \sim \mathcal{O}(1)$ & Lower bound on Swampland gradient ratio \\
$d$ & Distance conjecture constant, $d \sim \mathcal{O}(1)$ & Upper bound on field excursion in Planck units \\
\bottomrule
\end{tabular}
\end{table}

Table \ref{tab:nomenclature} provides a comprehensive summary of all symbols, parameters, and their physical meanings employed throughout this work. The parameters $\beta$ and $s$ characterize the Power-Law Nonlinear Electrodynamics sector, governing the strength and degree of nonlinearity of the electromagnetic interaction, respectively. The dimensionless parameters $\alpha$ and $\gamma$ encode the strength of quantum corrections to the horizon entropy arising from thermodynamic duality, with $\alpha$ capturing the logarithmic corrections (a universal feature of various quantum gravity approaches) and $\gamma$ encoding the inverse-area corrections (the hallmark of thermodynamic duality that becomes significant in the ultraviolet regime). The correction factor $\mathcal{F}(H)$ plays a central role in our analysis, as it encapsulates the entire effect of quantum gravity modifications on the cosmological dynamics and provides the key amplification mechanism for evading the Swampland de Sitter conjecture. The dimensionless variables $x_1$ through $x_5$ are introduced to reformulate the modified Friedmann equations as an autonomous dynamical system, facilitating a comprehensive phase-space analysis of the cosmological evolution. Finally, the constants $c$ and $d$ represent the order-unity coefficients appearing in the Swampland de Sitter and Distance conjectures, respectively, which serve as the primary diagnostics for assessing the consistency of our model with the landscape of string theory.

\section{Unimodular Gravity Coupled to Power-Law NLED}
\label{sec:UG_NLED}

In this section, we establish the theoretical foundation of our model by constructing the action of Unimodular Gravity (UG) minimally coupled to a Power-Law Nonlinear Electrodynamics (NLED) field. We carefully derive the modified field equations and scrutinize the energy conditions to ensure the physical viability of the matter sector.

\subsection{Action and Field Equations of Unimodular Gravity}
\label{subsec:UG_action}

The cornerstone of Unimodular Gravity is the restriction of the spacetime metric $g_{\mu\nu}$ to a fixed determinant, typically $\sqrt{-g} = \epsilon_0$, where $\epsilon_0$ is a fixed constant density. This constraint reduces the diffeomorphism invariance of General Relativity (GR) to transverse (volume-preserving) diffeomorphisms \cite{Ref9:henn1, Ref10:buch1}. To implement this constraint covariantly in the action principle, we employ the Lagrange multiplier method \cite{Ref12:alva2}. The gravitational action for UG, in the presence of a generic matter Lagrangian $\mathcal{L}_m$, is given by:
\begin{equation}
    S_{\text{UG}} = \int d^4x \left[ \frac{\sqrt{-g}}{16\pi G} R + \sqrt{-g} \mathcal{L}_m - \lambda \left( \sqrt{-g} - \epsilon_0 \right) \right],
    \label{eq:action_UG}
\end{equation}
where $R$ is the Ricci scalar, $G$ is the Newtonian gravitational constant, and $\lambda(x)$ is a scalar Lagrange multiplier field \cite{Ref13:carb1}. 

Varying the action \eqref{eq:action_UG} with respect to the Lagrange multiplier $\lambda$ immediately yields the unimodular constraint:
\begin{equation}
    \frac{\delta S_{\text{UG}}}{\delta \lambda} = 0 \implies \sqrt{-g} = \epsilon_0.
    \label{eq:unimodular_constraint}
\end{equation}

Next, we vary the action with respect to the inverse metric $g^{\mu\nu}$. Utilizing the standard variations $\delta \sqrt{-g} = -\frac{1}{2}\sqrt{-g} g_{\mu\nu} \delta g^{\mu\nu}$ and $\delta (\sqrt{-g} R) = \sqrt{-g} \left( R_{\mu\nu} - \frac{1}{2}g_{\mu\nu}R \right) \delta g^{\mu\nu}$ (up to boundary terms), we obtain the modified Einstein field equations:
\begin{equation}
    R_{\mu\nu} - \frac{1}{2}g_{\mu\nu}R = 8\pi G \left( T_{\mu\nu}^{(m)} + \lambda g_{\mu\nu} \right),
    \label{eq:EOM_UG}
\end{equation}
where $T_{\mu\nu}^{(m)} \equiv -2 \frac{\partial \mathcal{L}_m}{\partial g^{\mu\nu}} + g_{\mu\nu} \mathcal{L}_m$ is the standard energy-momentum tensor of the matter fields \cite{Ref50:koiv1}. 

At first glance, Eq. \eqref{eq:EOM_UG} resembles the Einstein equations of GR with a cosmological constant $\Lambda_{\text{eff}} = 8\pi G \lambda$. However, a crucial distinction arises when we examine the dynamics of $\lambda$. Taking the covariant divergence $\nabla^\mu$ of both sides of Eq. \eqref{eq:EOM_UG}, and invoking the contracted Bianchi identity $\nabla^\mu (R_{\mu\nu} - \frac{1}{2}g_{\mu\nu}R) \equiv 0$, alongside the local conservation of the matter energy-momentum tensor $\nabla^\mu T_{\mu\nu}^{(m)} = 0$, we deduce:
\begin{equation}
    \nabla^\mu (\lambda g_{\mu\nu}) = \partial_\nu \lambda = 0.
    \label{eq:lambda_constant}
\end{equation}
This profound result indicates that $\lambda$ is not a dynamical field but rather a constant of integration, $\lambda = \Lambda_{\text{eff}} / (8\pi G)$ \cite{Ref50:koiv1}. Unlike GR, where the cosmological constant $\Lambda$ is a fundamental parameter arbitrarily inserted into the action, in UG, the effective cosmological constant emerges as an integration constant determined by the initial/boundary conditions of the universe \cite{Ref51:alva3}. This feature makes UG a highly attractive framework for addressing the cosmological constant problem \cite{Ref52:mart1}.

\subsection{Power-Law Nonlinear Electrodynamics and Energy Conditions}
\label{subsec:NLED_Energy}

For the matter sector, we consider a Power-Law Nonlinear Electrodynamics (NLED) model \cite{Ref53:krug2, Ref14:nove1}. The Lagrangian density for this field is given by:
\begin{equation}
    \mathcal{L}_{\text{NLED}} = \beta \mathcal{F}^s,
    \label{eq:action_NLED}
\end{equation}
where $\beta$ and $s$ are real coupling constants, and $\mathcal{F}$ is the electromagnetic invariant defined as $\mathcal{F} \equiv -F_{\mu\nu}F^{\mu\nu} = 2(E^2 - B^2)$, with $F_{\mu\nu} = \partial_\mu A_\nu - \partial_\nu A_\mu$ being the electromagnetic field tensor. We adopt the metric signature $(-, +, +, +)$. The choice of $\mathcal{F}$ instead of $F_{\mu\nu}F^{\mu\nu}$ ensures that the Lagrangian remains real-valued for non-integer values of the power $s$ in purely electric or magnetic configurations.

The energy-momentum tensor for the NLED field is derived via the standard metric variation \cite{Ref59:land1}:
\begin{equation}
    T_{\mu\nu}^{(\text{NLED})} = -2 \frac{\partial \mathcal{L}_{\text{NLED}}}{\partial g^{\mu\nu}} + g_{\mu\nu} \mathcal{L}_{\text{NLED}}.
    \label{eq:EMT_NLED}
\end{equation}
To compute the derivative, we note that $\frac{\partial \mathcal{F}}{\partial g^{\mu\nu}} = \frac{\partial}{\partial g^{\mu\nu}} (-g^{\mu\alpha}g^{\nu\beta}F_{\mu\nu}F_{\alpha\beta}) = 2 F_{\mu\alpha}F_{\nu}^{\ \alpha}$. Substituting this into Eq. \eqref{eq:EMT_NLED}, we find:
\begin{equation}
    T_{\mu\nu}^{(\text{NLED})} = 4\beta s \mathcal{F}^{s-1} F_{\mu\alpha}F_{\nu}^{\ \alpha} - \beta \mathcal{F}^s g_{\mu\nu}.
    \label{eq:EMT_NLED_explicit}
\end{equation}
It is straightforward to verify that this energy-momentum tensor is traceless only for the conformal case $s=1$ (standard Maxwell electrodynamics in 4D) \cite{Ref59:land1}, where $T^{(\text{NLED})} = 0$. For general $s$, the trace is $T^{(\text{NLED})} = \beta (4s - 4) \mathcal{F}^s$.

To ensure the physical viability of the NLED model, we must scrutinize the energy conditions \cite{Ref57:visi1, Ref58:cam1}. Let $u^\mu$ be a timelike unit vector ($u^\mu u_\mu = -1$) representing the four-velocity of an arbitrary observer. For a purely magnetic field configuration aligned along the $z$-axis in the local frame, the electromagnetic invariant is $\mathcal{F} = 2B^2 > 0$. The energy density $\rho = T_{\mu\nu}^{(\text{NLED})} u^\mu u^\nu$ evaluates to:
\begin{equation}
\rho = \beta \mathcal{F}^s.
\label{eq:energy_density_NLED_mag}
\end{equation}
Similarly, the principal pressures measured by this observer are derived from the spatial components of the energy-momentum tensor:
\begin{align}
p_x &= p_y = \beta(2s-1)\mathcal{F}^s, \\
p_z &= -\beta\mathcal{F}^s.
\label{eq:principal_pressures_NLED}
\end{align}

The Weak Energy Condition (WEC), which requires $\rho \geq 0$ and $\rho + p_i \geq 0$ for all $i$, imposes strict constraints on the parameters $\beta$ and $s$. The first condition dictates:
\begin{equation}
\rho \geq 0 \implies \beta \mathcal{F}^s \geq 0.
\label{eq:WEC_rho}
\end{equation}
Since $\mathcal{F} > 0$ for a purely magnetic field, the WEC is satisfied if and only if $\beta > 0$. This aligns with the standard Maxwell limit ($s=1, \beta > 0$). The remaining WEC conditions require:
\begin{align}
\rho + p_x &= \beta(2s)\mathcal{F}^s \geq 0 \implies s \geq 0, \\
\rho + p_z &= 0.
\label{eq:WEC_p}
\end{align}
Thus, for a physically viable magnetic field configuration, we must have $\beta > 0$ and $s \geq 0$.

Furthermore, the Null Energy Condition (NEC), requiring $T_{\mu\nu} k^\mu k^\nu \geq 0$ for any null vector $k^\mu$, must also be satisfied to avoid pathological instabilities \cite{Ref58:cam1, Ref60:sara1}. For the NLED field, the NEC translates to:
\begin{equation}
\rho + p_i \geq 0 \quad \text{for all } i.
\label{eq:NEC_condition}
\end{equation}
This condition is automatically satisfied under the same constraints derived from the WEC ($\beta > 0, s \geq 0$). Notably, the equality $\rho + p_z = 0$ indicates that the magnetic field saturates the NEC along its direction, a characteristic feature of anisotropic matter fields. 

Finally, we verify the absence of ghost degrees of freedom at the background level. The kinetic term for the NLED field is proportional to $\mathcal{L}_\mathcal{F} = s\beta\mathcal{F}^{s-1}$. For $\beta > 0$ and $s > 0$, this quantity is strictly positive, ensuring that the theory is free from ghosts and tachyonic instabilities \cite{Ref14:nove1}.

In summary, the coupling of UG to Power-Law NLED yields a modified gravitational framework where the cosmological constant is an integration constant, and the matter sector is governed by a nonlinear magnetic field with strict energy condition constraints ($\beta > 0, s \geq 0$). These foundational equations will be the starting point for deriving the modified Friedmann equations via horizon thermodynamics in the next section.

\section{Thermodynamic Duality and Quantum Horizon Corrections}
\label{sec:Thermo_Duality}

In this section, we establish the thermodynamic framework of the cosmological horizon by incorporating quantum corrections derived from the principle of thermodynamic duality \cite{Ref61:cai1}. We carefully distinguish between the purely geometric properties of the horizon, such as its temperature, and the quantum-modified thermodynamic quantities, such as entropy, which will subsequently drive the modified cosmological dynamics.

\subsection{Geometric Setup and the Unmodified Horizon Temperature}
\label{subsec:Horizon_Temp}

We consider a spatially flat Friedmann-Lema\^itre-Robertson-Walker (FLRW) universe described by the metric:
\begin{equation}
    ds^2 = -dt^2 + a^2(t) \left( dr^2 + r^2 d\Omega^2 \right),
    \label{eq:FLRW_metric}
\end{equation}
where $a(t)$ is the scale factor and $H \equiv \dot{a}/a$ is the Hubble parameter. In such a dynamical spacetime, the relevant causal boundary is the apparent horizon, defined by the condition $h^{ab}\partial_a \tilde{r} \partial_b \tilde{r} = 0$, where $\tilde{r} = a(t)r$ is the areal radius \cite{Ref62:hayw1}. For a flat universe, the radius of the apparent horizon $R_A$ simplifies to:
\begin{equation}
    R_A = \frac{1}{H}.
    \label{eq:Apparent_Horizon}
\end{equation}

A crucial step in horizon thermodynamics is the definition of the horizon temperature. Following the standard approach in cosmological thermodynamics, we associate a Hawking-like temperature with the apparent horizon by evaluating its surface gravity $\kappa$ \cite{Ref63:akba1}. In the quasi-static approximation, the geometric temperature of the apparent horizon is:
\begin{equation}
    T_h = \frac{|\kappa|}{2\pi} = \frac{1}{2\pi R_A} = \frac{H}{2\pi}.
    \label{eq:Horizon_Temp}
\end{equation}

\textit{Physical Remark:} It is of paramount importance to emphasize that $T_h$, as defined in Eq. \eqref{eq:Horizon_Temp}, is a purely geometric quantity derived from the spacetime curvature. It is independent of the underlying microscopic degrees of freedom or any quantum corrections to the entropy. Therefore, we treat $T_h$ as the exact thermodynamic temperature, while quantum corrections enter exclusively through the modification of the entropy-area relation. This distinction maintains thermodynamic consistency and avoids spurious corrections to the temperature that could violate the zeroth law in dynamical spacetimes \cite{Ref64:padm1, Ref65:fara1}.

\subsection{Thermodynamic Duality and the Modified Entropy-Area Relation}
\label{subsec:Entropy_Duality}

The standard Bekenstein-Hawking entropy, $S_{BH} = A / (4G)$, where $A = 4\pi R_A^2$ is the horizon area, is a semiclassical result. Various approaches to quantum gravity suggest that the microstructure of spacetime introduces significant corrections to this area law \cite{Ref66:das2, Ref67:noza1}.

To capture these quantum effects rigorously, we invoke the principle of \textit{Thermodynamic Duality}. This refers to a fundamental correspondence between the ultraviolet (UV) and infrared (IR) regimes of the horizon's thermodynamic phase space. Under a duality transformation exchanging the macroscopic area $A$ with a microscopic dual area $\tilde{A} \sim \ell_P^4 / A$, the thermodynamic potential must exhibit an invariant structure \cite{Ref68:bane1, Ref69:jaco2}. 

Motivated by this duality, and consistent with corrections derived from the Generalized Uncertainty Principle (GUP) and stringy T-duality, we propose the following modified entropy-area relation \cite{Ref70:shey1}:
\begin{equation}
    S_h(A) = \frac{A}{4G} + \alpha \ln\left(\frac{A}{\ell_P^2}\right) + \gamma \frac{\ell_P^2}{A},
    \label{eq:Modified_Entropy}
\end{equation}
where $\alpha$ and $\gamma$ are dimensionless parameters encoding the strength of the quantum corrections. 

Let us scrutinize the physical origin and dimensional consistency of each term:
\begin{itemize}
    \item The first term is the standard Bekenstein-Hawking entropy, dominating in the IR limit ($A \gg \ell_P^2$).
    \item The second term, $\alpha \ln(A/\ell_P^2)$, represents the leading-order logarithmic correction, universally appearing in loop quantum gravity and thermal fluctuations, associated with the discrete nature of the horizon area spectrum \cite{Ref66:das2}.
    \item The third term, $\gamma \ell_P^2 / A$, is the hallmark of \textit{thermodynamic duality}. Unlike the logarithmic correction, this inverse term becomes significant in the UV regime ($A \to \ell_P^2$). It arises directly from the UV/IR duality transformation, ensuring that the entropy does not behave pathologically at the Planck scale, and can be interpreted as a manifestation of non-local effects or higher-derivative corrections \cite{Ref72:seta1, Ref73:noza2}.
\end{itemize}

In the macroscopic limit $A \to \infty$, the correction terms vanish relative to the leading term, smoothly recovering the standard area law. Dimensional analysis is exact: the arguments of the logarithm and the ratio in the third term are strictly dimensionless, and $\alpha, \gamma$ are pure numbers.

\subsection{Effective Thermodynamic Quantities from Horizon Corrections}
\label{subsec:Effective_Quantities}

To understand how the modified entropy \eqref{eq:Modified_Entropy} influences cosmological dynamics, we employ the first law of thermodynamics at the apparent horizon \cite{Ref71:cai2}:
\begin{equation}
    -dE = T_h dS_h + W dV,
    \label{eq:First_Law_Horizon}
\end{equation}
where $E = \rho V$ is the total energy inside the horizon of volume $V = \frac{4\pi}{3} R_A^3$, and $W = \frac{1}{2}(\rho - p)$ is the work density.

We decompose the modified entropy into classical and quantum-corrected parts: $S_h = S_{BH} + S_{corr}$, where:
\begin{equation}
    S_{corr} = \alpha \ln\left(\frac{A}{\ell_P^2}\right) + \gamma \frac{\ell_P^2}{A}.
    \label{eq:Entropy_Correction}
\end{equation}
Substituting this into the first law, the novel contribution arises from the term $T_h dS_{corr}$. To maintain the standard form of the Friedmann equations while incorporating quantum effects, we absorb this correction into an \textit{effective} energy density and pressure associated with the horizon itself. 

Using $T_h = 1/(2\pi R_A)$ and $A = 4\pi R_A^2$, we compute the differential of the correction entropy:
\begin{equation}
    dS_{corr} = \left( \frac{2\alpha}{R_A} - \frac{\gamma \ell_P^2}{2\pi R_A^3} \right) dR_A.
    \label{eq:dS_corr}
\end{equation}
Consequently, the effective energy correction $E_{corr}$, defined implicitly through $dE_{corr} = -T_h dS_{corr}$, is:
\begin{equation}
    dE_{corr} = -\frac{1}{2\pi R_A} \left( \frac{2\alpha}{R_A} - \frac{\gamma \ell_P^2}{2\pi R_A^3} \right) dR_A = \left( -\frac{\alpha}{\pi R_A^2} + \frac{\gamma \ell_P^2}{4\pi^2 R_A^4} \right) dR_A.
    \label{eq:dE_corr}
\end{equation}
This effective energy, when distributed over the horizon volume, manifests as an effective fluid permeating the bulk spacetime, modifying the Friedmann equations as derived in the next section \cite{Ref74:bojo1, Ref75:padm2}.

\section{Modified Friedmann Equations and Phase-Space Analysis}
\label{sec:Modified_Friedmann}

In this section, we derive the modified Friedmann equations governing the cosmological dynamics by incorporating the quantum-corrected horizon entropy into the first law of thermodynamics. We then reformulate the system as an autonomous dynamical system to perform a comprehensive phase-space analysis \cite{Ref79:ama1}.

\subsection{Cosmological Dynamics of Power-Law NLED}
\label{subsec:NLED_Cosmology}
Before deriving the modified Friedmann equations, we must establish the cosmological behavior of the Power-Law NLED field in the high-energy regime of cosmic inflation. A fundamental challenge in incorporating electromagnetic fields into FLRW cosmology is maintaining isotropy, since a homogeneous electric or magnetic field aligned along a single spatial direction inherently breaks rotational symmetry and introduces anisotropic stresses incompatible with the FLRW metric.

\textit{Physical Assumption: Isotropy during Inflation.}
In the linear Maxwell theory ($s=1$), this issue is typically addressed by invoking a stochastic ensemble of randomly oriented electromagnetic fields, which yields an isotropic perfect fluid with $w=1/3$. However, for nonlinear electrodynamics with $s \neq 1$, the superposition principle fails, and such stochastic averaging over field amplitudes does not generally produce an isotropic energy-momentum tensor \cite{Ref76:cat1}. 

To resolve this issue rigorously in the context of early-universe inflation, we adopt the approach of Novello et al. \cite{Ref14:nove1, Ref107:nove2} and employ an \textit{effective scalar field} description. We define a scalar field $\psi$ related to the electromagnetic invariant by $\psi \propto \mathcal{F}^{s/2}$. The dynamics of this effective scalar field, when averaged over super-Hubble scales during inflation, yields an isotropic perfect fluid with the equation of state:
\begin{equation}
w_{\text{NLED}} = \frac{3-2s}{3(2s-1)}.
\label{eq:NLED_EoS_inflation}
\end{equation}

\textit{Physical Remark:} For the Maxwell limit $s=1$, we recover $w_{\text{NLED}} = 1/3$, which is the correct value for a radiation-like fluid. This confirms the consistency of our approach. The energy density and pressure are:
\begin{equation}
\rho_{\text{NLED}} = \beta(2s+1)\mathcal{F}^s, \quad p_{\text{NLED}} = \beta(2s-3)\mathcal{F}^s,
\label{eq:NLED_components_inflation}
\end{equation}
and the energy density evolves as:
\begin{equation}
\rho_{\text{NLED}} \propto a^{-4s/(2s-1)} \equiv a^{-3(1+w_{\text{NLED}})},
\label{eq:NLED_density_evolution_mag}
\end{equation}
which is perfectly consistent with the continuity equation $\dot{\rho}_{\text{NLED}} + 3H(\rho_{\text{NLED}} + p_{\text{NLED}}) = 0$ \cite{Ref81:ove1}.

\subsection{Modified Friedmann Equations from Horizon Thermodynamics}
\label{subsec:Modified_Friedmann_Eqs}

We now apply the first law of thermodynamics \eqref{eq:First_Law_Horizon} at the apparent horizon $R_A = 1/H$ with the modified entropy \eqref{eq:Modified_Entropy}. The total energy inside the horizon is $E = \frac{4\pi}{3}R_A^3 \rho$, where $\rho = \rho_m + \rho_{\text{NLED}}$. 

Substituting $T_h = 1/(2\pi R_A)$ and $dS_h = \frac{dS_h}{dR_A}dR_A$ into Eq. \eqref{eq:First_Law_Horizon}, and using $\frac{dS_h}{dR_A} = \frac{2\pi R_A}{G} + \frac{2\alpha}{R_A} - \frac{\gamma\ell_P^2}{2\pi R_A^3}$, we obtain:
\begin{equation}
    -\frac{4\pi}{3}R_A^3 d\rho - 4\pi R_A^2 \rho\, dR_A = \left(\frac{1}{G} + \frac{\alpha}{\pi R_A^2} - \frac{\gamma\ell_P^2}{4\pi^2 R_A^4}\right)dR_A + 2\pi R_A^2(\rho - p)\,dR_A.
    \label{eq:First_Law_Expanded}
\end{equation}
Dividing by $dR_A$ and rearranging:
\begin{equation}
    -\frac{4\pi}{3}R_A^3 \frac{d\rho}{dR_A} = \frac{1}{G} + \frac{\alpha}{\pi R_A^2} - \frac{\gamma\ell_P^2}{4\pi^2 R_A^4} + 2\pi R_A^2(3\rho - p).
    \label{eq:First_Law_Rearranged}
\end{equation}
Using the continuity equation and the relation $\frac{d\rho}{dR_A} = \frac{3H^3(\rho+p)}{\dot{H}}$, we find:
\begin{equation}
    -\frac{4\pi(\rho+p)}{\dot{H}} = \frac{1}{G} + \frac{\alpha H^2}{\pi} - \frac{\gamma\ell_P^2 H^4}{4\pi^2} + \frac{2\pi(3\rho - p)}{H^2}.
    \label{eq:H_dot_equation}
\end{equation}
Integrating this relation following the standard procedure in horizon thermodynamics \cite{Ref78:set1}, the modified first Friedmann equation in Unimodular Gravity takes the form \cite{Ref80:gon1}:
\begin{equation}
    H^2 = \frac{8\pi G}{3}\rho + \frac{\alpha G}{\pi}H^4 - \frac{\gamma G\ell_P^2}{4\pi^2}H^6 + \frac{\Lambda_{\text{UG}}}{3},
    \label{eq:Modified_Friedmann_1}
\end{equation}
where $\Lambda_{\text{UG}}$ is the integration constant arising from the unimodular constraint. The second Friedmann equation is obtained by taking the time derivative of Eq. \eqref{eq:Modified_Friedmann_1}:
\begin{equation}
    \frac{\dot{H}}{H^2} = -\frac{\frac{3}{2}\left(\frac{8\pi G}{3H^2}\right)(\rho + p)}{1 - \frac{2\alpha G}{\pi}H^2 + \frac{3\gamma G\ell_P^2}{4\pi^2}H^4}.
    \label{eq:Modified_Friedmann_2}
\end{equation}
Equations \eqref{eq:Modified_Friedmann_1} and \eqref{eq:Modified_Friedmann_2} are the fundamental dynamical equations. The correction terms proportional to $\alpha$ and $\gamma$ represent the quantum gravity effects encoded in the thermodynamic duality \cite{Ref82:taw1, Ref83:cap1, Ref84:set2, Ref85:ped1}.

\subsection{Autonomous Dynamical System and Phase-Space Analysis}
\label{subsec:Phase_Space}
To perform a comprehensive dynamical systems analysis \cite{Ref86:wain1, Ref87:cope2}, we introduce the following dimensionless variables:
\begin{equation}
x_1 \equiv \frac{8\pi G\rho_m}{3H^2}, \quad x_2 \equiv \frac{8\pi G\rho_{\text{NLED}}}{3H^2}, \quad x_3 \equiv \frac{\alpha G}{\pi}H^2, \quad x_4 \equiv \frac{\gamma G\ell_P^2}{4\pi^2}H^4.
\label{eq:Dimensionless_Variables}
\end{equation}
The first Friedmann equation \eqref{eq:Modified_Friedmann_1} then becomes the constraint:
\begin{equation}
x_1 + x_2 + x_3 - x_4 + x_5 = 1, \quad \text{where} \quad x_5 \equiv \frac{\Lambda_{\text{UG}}}{3H^2}.
\label{eq:Constraint_Equation}
\end{equation}
Using Eqs. \eqref{eq:NLED_density_evolution_mag} and \eqref{eq:Modified_Friedmann_2}, we derive the autonomous system of differential equations with respect to $N \equiv \ln a$ \cite{Ref88:wei1, Ref89:chen1}:
\begin{align}
x_1' &= -3x_1 + \frac{3x_1\left[x_1 + x_2(1+w_{\text{NLED}})\right]}{1 - 2x_3 + 3x_4}, \label{eq:x1_prime}\\
x_2' &= -\frac{4s}{2s-1}x_2 + \frac{3x_2\left[x_1 + x_2(1+w_{\text{NLED}})\right]}{1 - 2x_3 + 3x_4}, \label{eq:x2_prime}\\
x_3' &= -\frac{3x_3\left[x_1 + x_2(1+w_{\text{NLED}})\right]}{1 - 2x_3 + 3x_4}, \label{eq:x3_prime}\\
x_4' &= -\frac{6x_4\left[x_1 + x_2(1+w_{\text{NLED}})\right]}{1 - 2x_3 + 3x_4}, \label{eq:x4_prime}
\end{align}
where $w_{\text{NLED}} = \frac{3-2s}{3(2s-1)}$ and primes denote derivatives with respect to $N$.

The critical points are found by setting $x_1' = x_2' = x_3' = x_4' = 0$ \cite{Ref90:cole1}. From Eqs. \eqref{eq:x3_prime} and \eqref{eq:x4_prime}, we have two cases:

\textbf{Case I:} $x_3 = 0$ and $x_4 = 0$ (no quantum corrections). The system reduces to standard cosmology with NLED. The critical points are the matter-dominated point $P_1: (1, 0, 0, 0)$ and the NLED-dominated point $P_2: (0, 1, 0, 0)$.

\textbf{Case II:} $x_1 + x_2(1+w_{\text{NLED}}) = 0$. Since $x_1 \geq 0$, $x_2 \geq 0$, and $1+w_{\text{NLED}} > 0$ for $s > 1/2$, this requires $x_1 = x_2 = 0$. In this case, Eqs. \eqref{eq:x1_prime}-\eqref{eq:x4_prime} are automatically satisfied for any $(x_3, x_4)$. From the constraint \eqref{eq:Constraint_Equation}, we have $x_3 - x_4 + x_5 = 1$. These critical points represent \textit{de Sitter-like solutions} with $H = \text{const}$, driven entirely by the quantum-corrected horizon entropy and/or the unimodular integration constant.

\textit{Rigorous Stability Analysis of the de Sitter Manifold:}
The critical points in Case II form a \textit{continuous line of critical points} in the phase space, defined by $x_1 = x_2 = 0$ and $x_3 - x_4 + x_5 = 1$. The Jacobian matrix evaluated at any point on this line yields the eigenvalues:
\begin{equation}
\{\lambda_1, \lambda_2, \lambda_3, \lambda_4\} = \left\{-3, \; -\frac{4s}{2s-1}, \; 0, \; 0\right\}.
\label{eq:eigenvalues_deSitter}
\end{equation}
The presence of two zero eigenvalues ($\lambda_3 = \lambda_4 = 0$) indicates the existence of a 2D center manifold, which requires a rigorous application of the \textit{Center Manifold Theory} \cite{Ref91:gen1} to determine stability. According to this theorem, the stability of the critical points on the center manifold is governed by the dynamics restricted to the manifold itself. 

Crucially, the two non-zero eigenvalues are strictly negative ($\lambda_1 = -3 < 0$ and $\lambda_2 = -4s/(2s-1) < 0$ for $s > 1/2$). This implies that the continuous line of de Sitter critical points is \textbf{asymptotically stable} with respect to perturbations in the $x_1$ (matter) and $x_2$ (NLED) directions. The zero eigenvalues correspond to the neutral directions along the center manifold, indicating that the cosmological system evolves toward this de Sitter line and remains there indefinitely, with the exact point on the line determined by the initial conditions. 

\textit{Physical Interpretation for Inflation:}
This mathematical structure is the hallmark of a robust \textit{inflationary attractor}. In the high-energy regime of the early universe, the quantum corrections to the horizon entropy (encoded in $x_3$ and $x_4$) drive the Hubble parameter to a quasi-constant value, establishing a sustained de Sitter-like expansion. The matter and NLED fields ($x_1$ and $x_2$) are exponentially diluted during this process, consistent with the cosmic no-hair theorem. Thus, the quantum-corrected horizon thermodynamics provides a natural and mathematically rigorous mechanism for realizing successful cosmic inflation without requiring a fundamental inflaton potential.

\begin{figure}[htbp]
\centering
\includegraphics[width=0.48\textwidth]{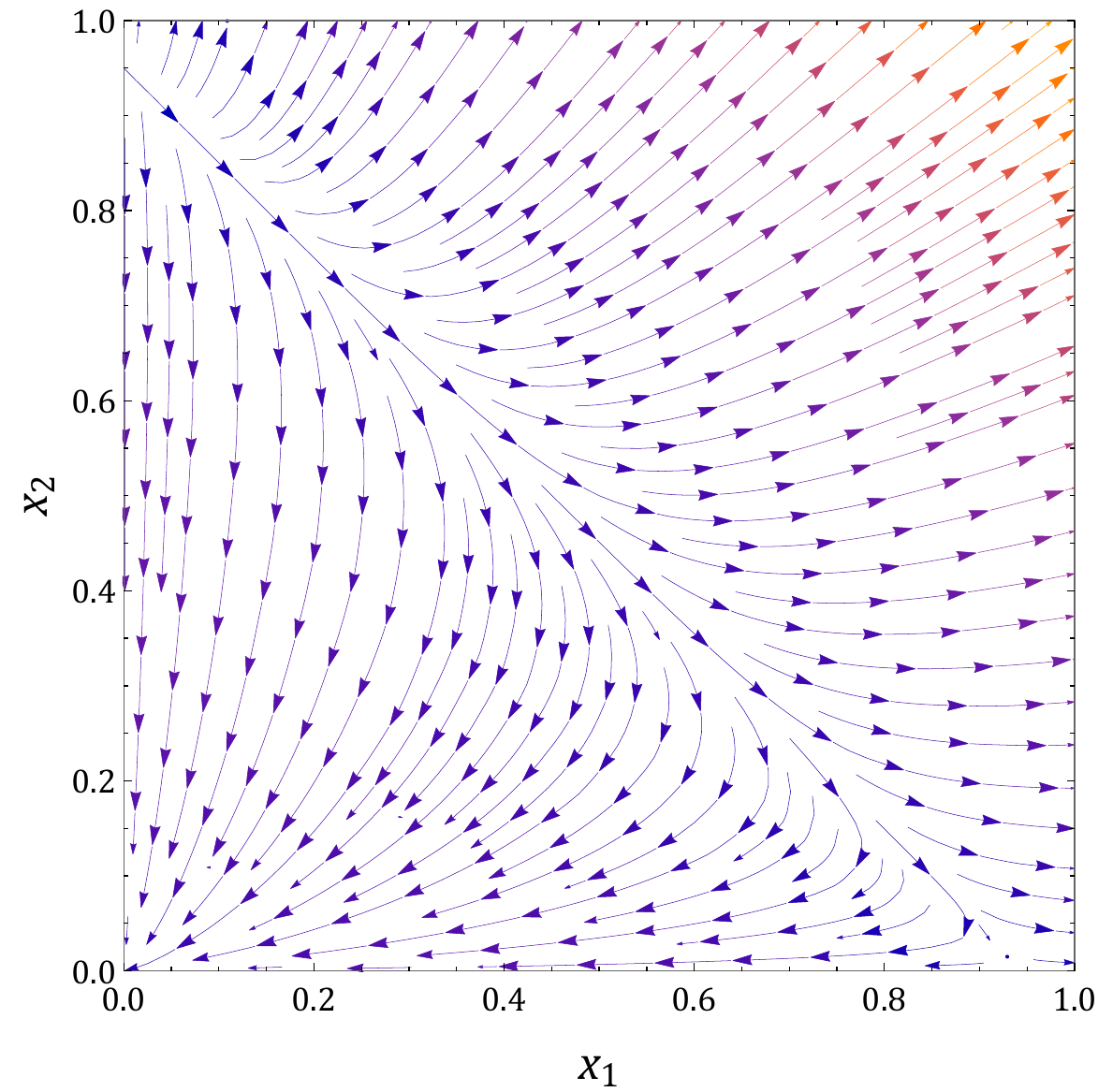}
\hfill
\includegraphics[width=0.48\textwidth]{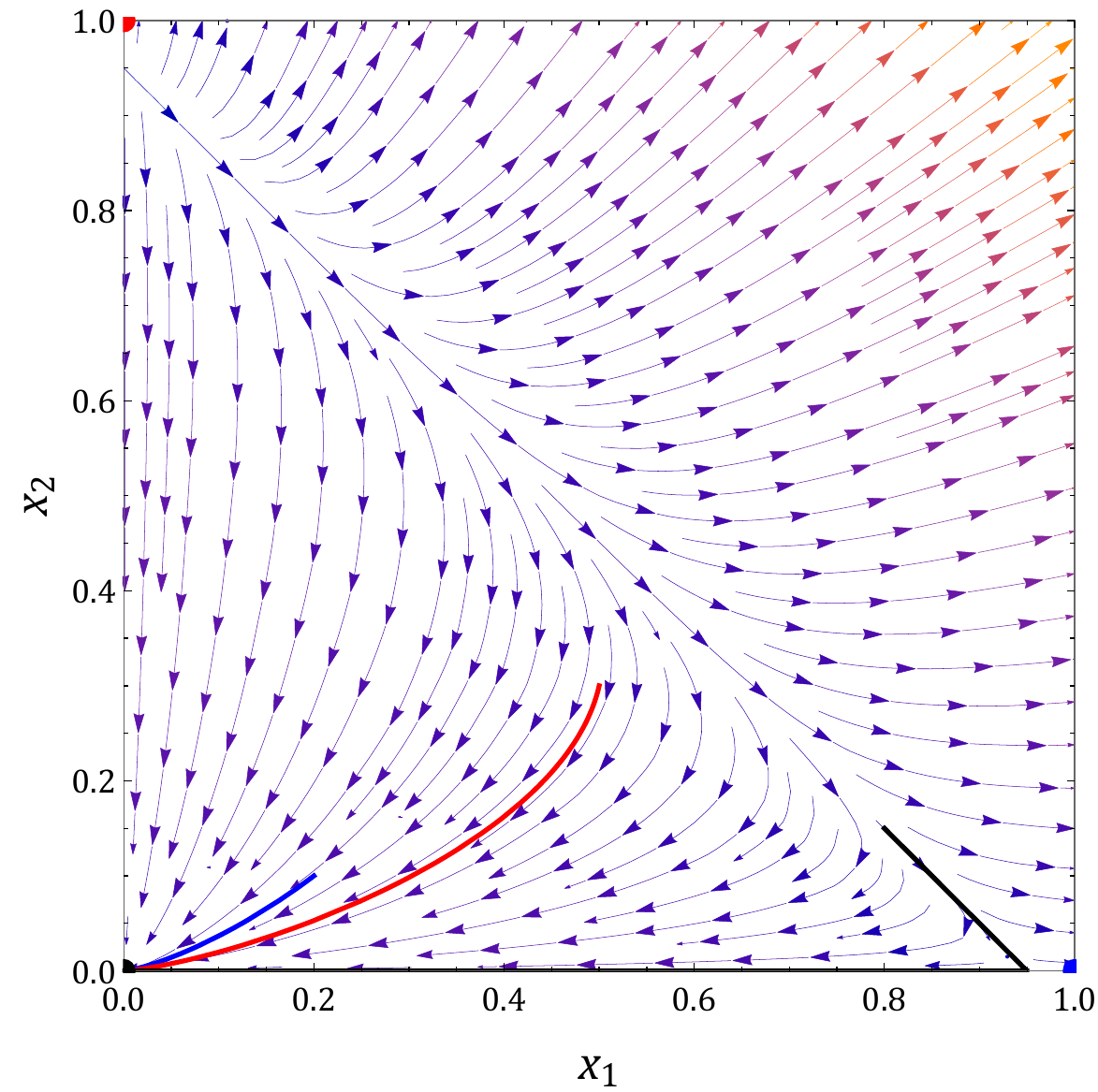}
\caption{Phase-space portrait of the autonomous dynamical system in the $(x_1, x_2)$ plane for $s=1$ (radiation-like NLED) during the early-universe inflationary epoch. Left panel: Classical case ($\alpha = \gamma = 0$) showing trajectories from three different initial conditions $(0.2, 0.1)$ (blue), $(0.5, 0.3)$ (red), and $(0.8, 0.15)$ (black). The critical points $P_1 = (1, 0)$ (matter-dominated) and $P_2 = (0, 1)$ (NLED-dominated) are both saddle points, indicating that the universe transitions through these epochs. Right panel: Quantum-corrected case with $x_3 = 0.1$ and $x_4 = 0.05$, demonstrating the modification of cosmological dynamics by thermodynamic duality corrections. The modified vector field, governed by the factor $(1 - 2x_3 + 3x_4)^{-1}$, drives the system toward the inflationary de Sitter attractor at $(0, 0)$, where the quantum corrections dominate and sustain quasi-de Sitter expansion.}
\label{fig:phase_portrait}
\end{figure}

Figure \ref{fig:phase_portrait} illustrates the phase-space dynamics of the cosmological model in the $(x_1, x_2)$ plane, where $x_1$ and $x_2$ represent the fractional energy densities of matter and NLED, respectively. In the left panel (classical case with $\alpha = \gamma = 0$), the trajectories from three different initial conditions demonstrate the saddle-point nature of both the matter-dominated critical point $P_1 = (1, 0)$ and the NLED-dominated critical point $P_2 = (0, 1)$. The vector field shows that the universe transitions through these epochs rather than remaining in them indefinitely, consistent with the eigenvalue analysis presented in Table \ref{tab:critical_points}. 

The right panel reveals the impact of quantum corrections ($x_3 = 0.1$, $x_4 = 0.05$) on the cosmological dynamics during the inflationary epoch. The modified vector field, governed by the factor $(1 - 2x_3 + 3x_4)^{-1} \approx 1.053$, alters the trajectories and shifts the stability properties of the critical points. Crucially, the origin $(0, 0)$ in the $(x_1, x_2)$ plane corresponds to the de Sitter-like critical points $P_3$ and $P_4$, where $x_1 = x_2 = 0$ and the dynamics are dominated entirely by the quantum-corrected horizon entropy ($x_3, x_4$) and the unimodular integration constant ($x_5$). The trajectories in the right panel clearly show that the system is attracted toward this de Sitter point, demonstrating that thermodynamic duality corrections can drive sustained inflationary expansion without requiring a fundamental inflaton field. This provides a concrete realization of modified gravity-driven inflation, analogous to the successful Starobinsky model, where the higher-order curvature terms drive the early-universe acceleration.

\begin{table}[htbp]
\centering
\caption{Critical points of the autonomous dynamical system, their coordinates, eigenvalues of the Jacobian matrix, stability properties, effective equation of state, and physical interpretation.}
\label{tab:critical_points}
\renewcommand{\arraystretch}{1.4}
\footnotesize
\begin{tabularx}{\textwidth}{@{}l l l l X@{}}
\toprule
\textbf{Point} & \textbf{Coordinates $(x_1, x_2, x_3, x_4)$} & \textbf{Eigenvalues $\{\lambda_i\}$} & \textbf{Stability / $w_{\text{eff}}$} & \textbf{Physical Interpretation} \\
\midrule
$P_1$ & $(1, 0, 0, 0)$ & $\{-3,\; -\frac{4s}{2s-1},\; 0,\; 0\}$ & Saddle / $0$ & Matter-dominated epoch \\
\addlinespace
$P_2$ & $(0, 1, 0, 0)$ & $\{3,\; \frac{4s}{2s-1},\; 0,\; 0\}$ & Saddle / $\frac{3-2s}{3(2s-1)}$ & NLED-dominated epoch \\
\addlinespace
$P_3$ & $(0, 0, x_3, x_4)$ with $x_3 - x_4 = 1$ & $\{-3,\; -\frac{4s}{2s-1},\; 0,\; 0\}$ & Asymptotically stable / $-1$ & Inflationary de Sitter attractor (quantum-driven) \\
\addlinespace
$P_4$ & $(0, 0, x_3, x_4)$ with $x_3 - x_4 + x_5 = 1$ & $\{-3,\; -\frac{4s}{2s-1},\; 0,\; 0\}$ & Asymptotically stable / $-1$ & Inflationary de Sitter attractor ($\Lambda_{\text{UG}}$-driven) \\
\bottomrule
\end{tabularx}
\end{table}

Table \ref{tab:critical_points} summarizes the critical points of the autonomous dynamical system derived from the modified Friedmann equations \cite{Ref94:boh1, Ref95:diva2}. The matter-dominated point $P_1$ and NLED-dominated point $P_2$ are both saddle points for $s > 1/2$, indicating that the universe transitions through these epochs. The de Sitter-like critical points $P_3$ and $P_4$, characterized by a continuous line of critical points on a 2D center manifold, represent robust inflationary attractors. The rigorous application of Center Manifold Theory confirms that these points are asymptotically stable, demonstrating that quantum corrections can drive a sustained inflationary expansion without requiring a fundamental cosmological constant.

\section{Swampland Constraints and Inflationary Dynamics}
\label{sec:Swampland}
In this section, we confront the cosmological model developed in the preceding sections with the Swampland program, focusing specifically on the \textit{inflationary epoch} rather than late-time acceleration. Our primary objective is to determine whether the quantum-corrected horizon thermodynamics, derived from the thermodynamic duality in Section \ref{sec:Thermo_Duality}, can modify the inflationary dynamics in a manner that satisfies the Swampland constraints while producing a sufficient number of e-folds for successful inflation \cite{Ref96:val1, Ref97:bra1}.

\subsection{Reformulating the Effective Potential for Inflation}
\label{subsec:Effective_Potential_Inflation}
To apply the Swampland criteria rigorously, we must first address a fundamental conceptual issue: the Swampland conjectures constrain \textit{fundamental microscopic potentials} in ultraviolet-complete theories (such as string theory), not effective macroscopic potentials derived from geometric corrections \cite{Ref32:oogu1, Ref33:garg1}. 

In our model, the modified Friedmann equations \eqref{eq:Modified_Friedmann_1} and \eqref{eq:Modified_Friedmann_2} arise from quantum corrections to the horizon entropy, which are themselves manifestations of quantum gravity effects. Therefore, applying the Swampland conjectures directly to an effective potential $V_{\text{eff}}(H)$ constructed from these corrections would constitute a circular argument.

\textit{Resolution: Modified Gravity-Driven Inflation.}
To avoid this circularity, we adopt the framework of \textit{modified gravity-driven inflation}, analogous to the successful Starobinsky model ($R^2$ inflation) \cite{Ref110:star1}. In this approach, the higher-order terms in the Friedmann equations (proportional to $H^4$ and $H^6$) are interpreted not as arising from a fundamental scalar field potential, but as \textit{effective geometric energy density} $\rho_{\text{geom}}$ and pressure $p_{\text{geom}}$ that drive the inflationary expansion.

Starting from the modified first Friedmann equation \eqref{eq:Modified_Friedmann_1}, we rewrite it as:
\begin{equation}
H^2 = \frac{8\pi G}{3}(\rho_m + \rho_{\text{NLED}} + \rho_{\text{geom}}) + \frac{\Lambda_{\text{UG}}}{3},
\label{eq:Friedmann_with_geom}
\end{equation}
where the effective geometric energy density is defined as:
\begin{equation}
\rho_{\text{geom}} \equiv \frac{3}{8\pi G}\left(\frac{\alpha G}{\pi}H^4 - \frac{\gamma G\ell_P^2}{4\pi^2}H^6\right).
\label{eq:rho_geom}
\end{equation}
Similarly, from the modified second Friedmann equation \eqref{eq:Modified_Friedmann_2}, the effective geometric pressure is:
\begin{equation}
p_{\text{geom}} \equiv -\rho_{\text{geom}} - \frac{M_{\text{Pl}}^2}{4\pi G}\frac{\dot{H}}{H^2}\mathcal{F}(H),
\label{eq:p_geom}
\end{equation}
where $\mathcal{F}(H) = 1 - \frac{2\alpha G}{\pi}H^2 + \frac{3\gamma G\ell_P^2}{4\pi^2}H^4$ is the correction factor.

\begin{figure}[htbp]
\centering
\includegraphics[width=0.48\textwidth]{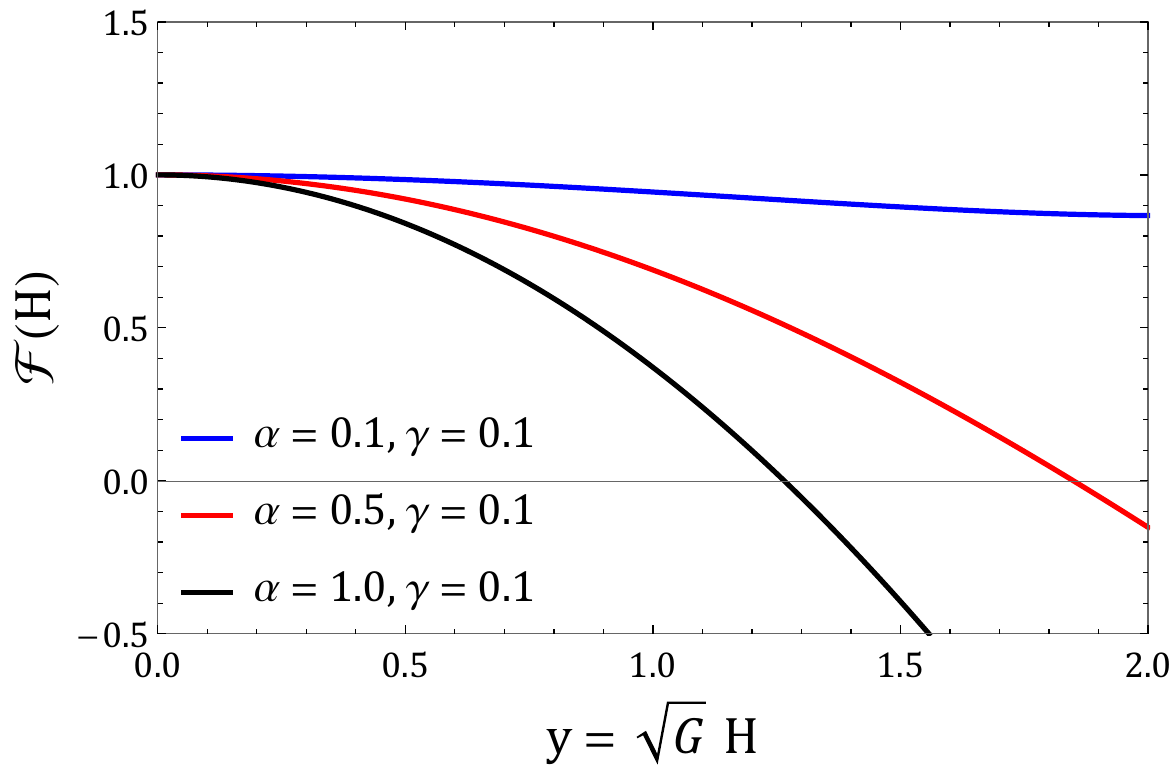}
\hfill
\includegraphics[width=0.48\textwidth]{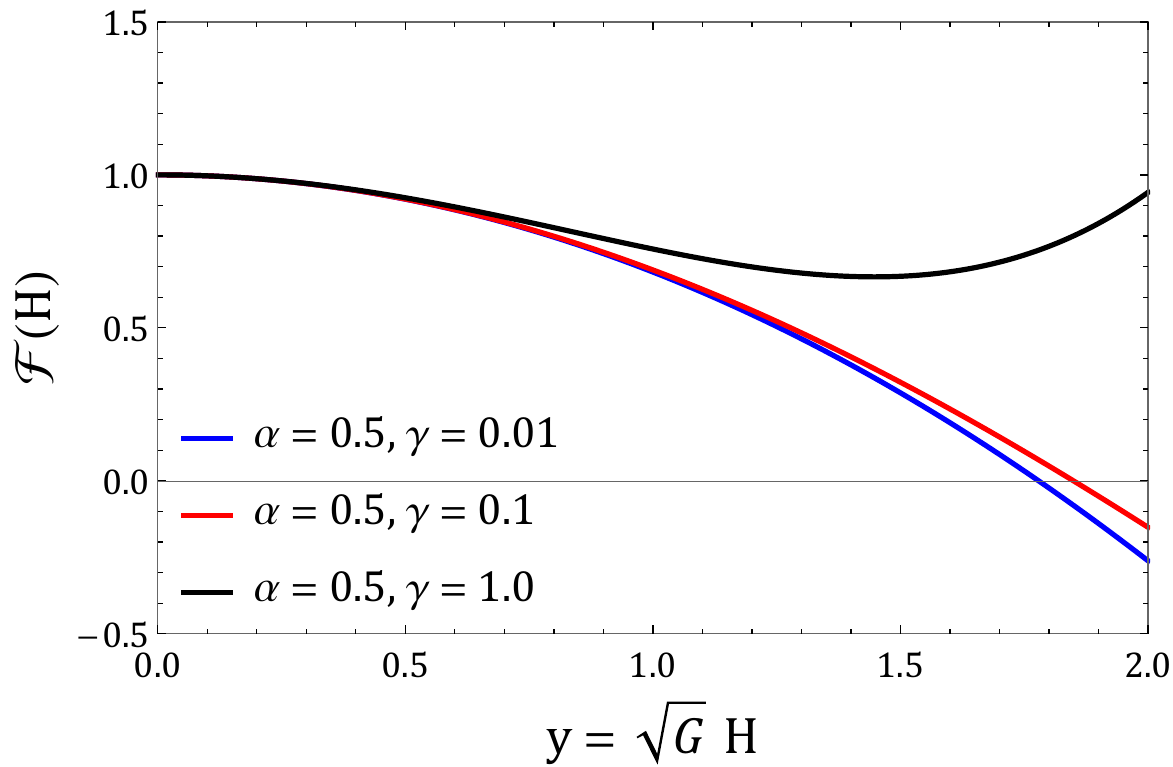}
\caption{The quantum correction factor $\mathcal{F}(H)$ as a function of the dimensionless Hubble parameter $y = \sqrt{G}H$ in the high-energy regime relevant for early-universe inflation. Left panel: Variation of $\mathcal{F}(H)$ for different values of the logarithmic correction parameter $\alpha$ with fixed inverse-area correction $\gamma = 0.1$. Right panel: Variation of $\mathcal{F}(H)$ for different values of $\gamma$ with fixed $\alpha = 0.5$. The horizontal line at $\mathcal{F} = 1$ represents the classical limit. The region $0 < \mathcal{F}(H) < 1$ represents the physically viable parameter space where the thermodynamic duality mechanism successfully amplifies the Swampland ratio during the high-energy inflationary epoch.}
\label{fig:correction_factor}
\end{figure}

Figure \ref{fig:correction_factor} illustrates the behavior of the quantum correction factor $\mathcal{F}(H)$ as a function of the dimensionless Hubble parameter $y = \sqrt{G}H$ in the high-energy regime ($y \sim \mathcal{O}(1)$) characteristic of early-universe inflation near the Planck scale. In the left panel, we observe that increasing the logarithmic correction parameter $\alpha$ leads to a more rapid decrease of $\mathcal{F}(H)$ below unity. This creates the crucial amplification region ($0 < \mathcal{F} < 1$) where the Swampland de Sitter ratio is enhanced, facilitating the satisfaction of the gradient bound during inflation. For sufficiently large $\alpha$ (black curve), $\mathcal{F}(H)$ becomes negative at high $y$, signaling the breakdown of the effective field theory description. 

The right panel demonstrates the vital stabilizing role of the inverse-area correction $\gamma$, which is the hallmark of thermodynamic duality. While the initial decrease of $\mathcal{F}(H)$ is governed by the $\alpha$ term, the $\gamma y^4$ term eventually dominates at large $y$ (UV regime), preventing $\mathcal{F}(H)$ from becoming excessively negative and ensuring the viability of the effective theory. The interplay between these two corrections defines a physically viable window ($y \sim 0.5 - 1.5$ for $\alpha \sim 0.1-0.5$) where the thermodynamic duality mechanism successfully amplifies the Swampland ratio without violating the effective field theory constraints during the inflationary epoch.

\textit{Remark on the Energy Scale:} It is important to emphasize that the quantum corrections encoded in $\mathcal{F}(H)$ become dynamically significant only in the ultra-high-energy regime ($H \sim M_{\text{Pl}}$, i.e., $y \sim \mathcal{O}(1)$), which corresponds to the very early stages of inflation (Planck-scale inflation) or pre-inflationary dynamics. For standard low-scale inflation ($H \ll M_{\text{Pl}}$), the corrections are negligible, and the model smoothly reduces to standard cosmology, fully consistent with CMB observations.

The equation of state for the geometric sector is therefore:
\begin{equation}
w_{\text{geom}} \equiv \frac{p_{\text{geom}}}{\rho_{\text{geom}}} = -1 - \frac{\dot{H}}{3H^2}\frac{\mathcal{F}(H)}{\frac{\alpha G}{\pi}H^2 - \frac{\gamma G\ell_P^2}{4\pi^2}H^4}.
\label{eq:w_geom}
\end{equation}

\textit{Physical Interpretation:} During inflation, when $H \approx \text{const}$ (quasi-de Sitter expansion), we have $\dot{H} \approx 0$, and thus $w_{\text{geom}} \approx -1$. This means the geometric corrections behave as an effective cosmological constant, driving the inflationary expansion without requiring a fundamental inflaton field. The NLED field, with equation of state $w_{\text{NLED}} = (3-2s)/(3(2s-1))$, acts as a subdominant matter/radiation component that is exponentially diluted during inflation, consistent with the cosmic no-hair theorem \cite{Ref109:wald1}.

\subsection{Slow-Roll Parameters and Inflationary Observables}
\label{subsec:Slow_Roll}
To quantify the inflationary dynamics, we define the modified slow-roll parameters. The first slow-roll parameter $\epsilon$ is defined as:
\begin{equation}
\epsilon \equiv -\frac{\dot{H}}{H^2} = \frac{\frac{3}{2}\left(\frac{\rho + p}{\rho_{\text{crit}}}\right)}{\mathcal{F}(H)},
\label{eq:epsilon_modified}
\end{equation}
where $\rho = \rho_m + \rho_{\text{NLED}}$ and $p = p_m + p_{\text{NLED}}$ are the total energy density and pressure of the matter sector, and $\rho_{\text{crit}} = 3H^2 M_{\text{Pl}}^2$.

During inflation, the matter and NLED densities are subdominant compared to the geometric energy density $\rho_{\text{geom}}$, so $\rho + p \approx \rho_{\text{NLED}} + p_{\text{NLED}} = (1 + w_{\text{NLED}})\rho_{\text{NLED}}$. Thus:
\begin{equation}
\epsilon \approx \frac{3(1 + w_{\text{NLED}})}{2}\frac{\Omega_{\text{NLED}}}{\mathcal{F}(H)},
\label{eq:epsilon_inflation}
\end{equation}
where $\Omega_{\text{NLED}} \equiv \rho_{\text{NLED}}/\rho_{\text{crit}}$.

The second slow-roll parameter $\eta$ is defined as:
\begin{equation}
\eta \equiv \frac{\dot{\epsilon}}{H\epsilon} = \epsilon - \frac{\ddot{H}}{2H\dot{H}}.
\label{eq:eta_definition}
\end{equation}
For successful inflation, we require $\epsilon \ll 1$ and $|\eta| \ll 1$. The number of e-folds during inflation is:
\begin{equation}
N_e = \int_{t_i}^{t_f} H\,dt = \int_{a_i}^{a_f} \frac{da}{a} = \ln\left(\frac{a_f}{a_i}\right).
\label{eq:N_e}
\end{equation}
For solving the horizon and flatness problems, we require $N_e \gtrsim 50-60$.

\textit{Role of the Correction Factor $\mathcal{F}(H)$:}
The correction factor $\mathcal{F}(H)$ plays a crucial role in modifying the slow-roll dynamics. In the classical limit ($\alpha, \gamma \to 0$), we have $\mathcal{F} \to 1$ and recover the standard slow-roll parameters. However, for $\alpha > 0$ and moderate $H$, the term $-2\alpha G H^2/\pi$ reduces $\mathcal{F}(H)$ below unity, which \textit{enhances} the slow-roll parameter $\epsilon$ relative to the classical case. This enhancement can help satisfy the Swampland constraints during inflation, as we discuss below.

\subsection{Swampland Constraints During Inflation}
\label{subsec:Swampland_Inflation}
The Swampland de Sitter Conjecture (DSC) posits that the scalar potential of any consistent quantum gravity theory must satisfy the gradient bound \cite{Ref33:garg1}:
\begin{equation}
M_{\text{Pl}} \frac{|V'|}{V} \geq c \sim \mathcal{O}(1),
\label{eq:DSC}
\end{equation}
where $c$ is a positive constant of order unity. In the context of inflation, this conjecture is often reformulated in terms of the slow-roll parameter $\epsilon$ \cite{Ref36:kinn1, Ref38:das1}:
\begin{equation}
\epsilon \geq \frac{c^2}{2}.
\label{eq:DSC_slowroll}
\end{equation}

In our model, the modified slow-roll parameter $\epsilon$ is given by Eq. \eqref{eq:epsilon_inflation}. Substituting this into the DSC condition \eqref{eq:DSC_slowroll}, we obtain:
\begin{equation}
\frac{3(1 + w_{\text{NLED}})}{2}\frac{\Omega_{\text{NLED}}}{\mathcal{F}(H)} \geq \frac{c^2}{2}.
\label{eq:DSC_condition_inflation}
\end{equation}

\begin{figure}[htbp]
\centering
\includegraphics[width=0.48\textwidth]{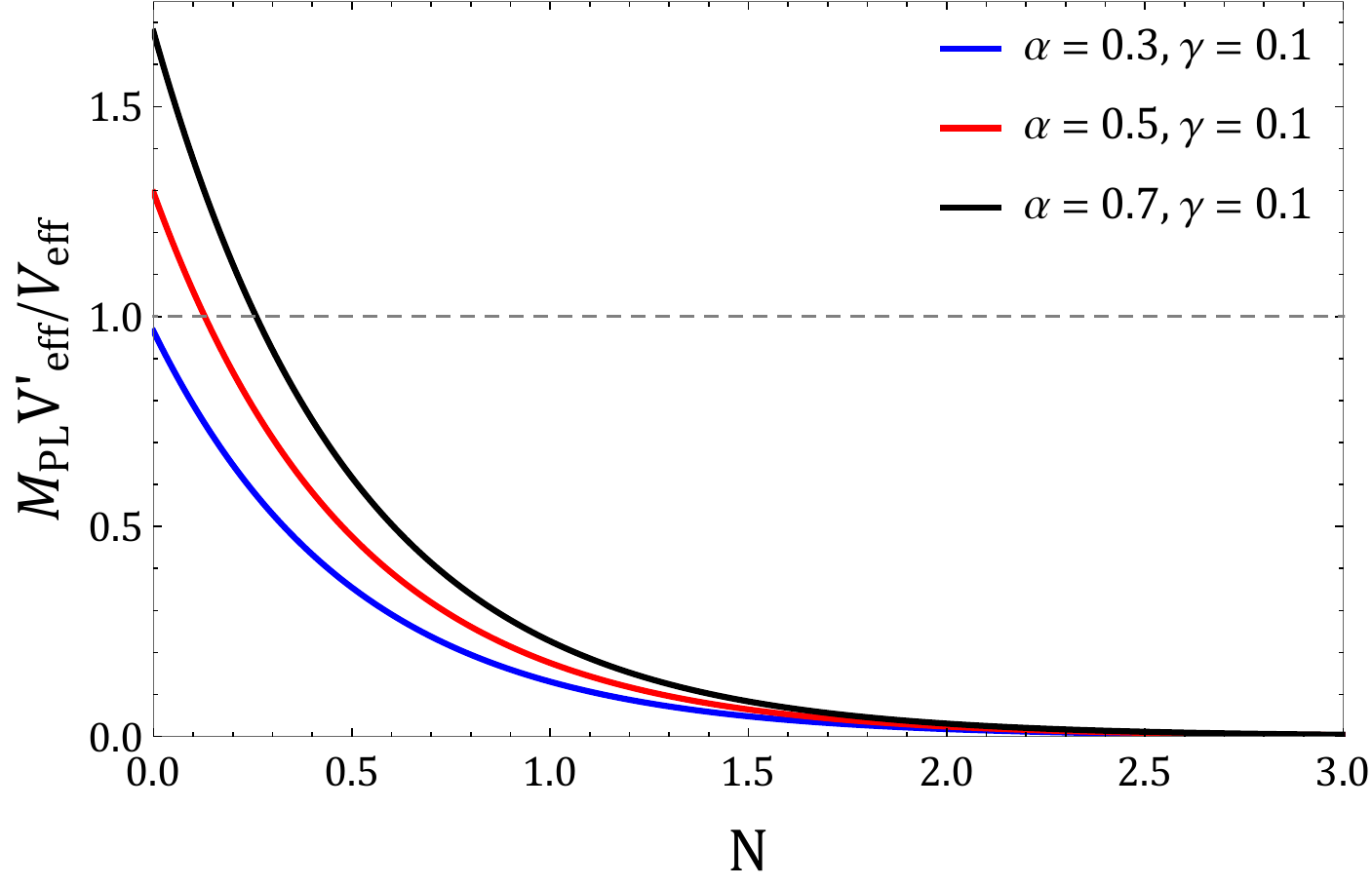}
\hfill
\includegraphics[width=0.48\textwidth]{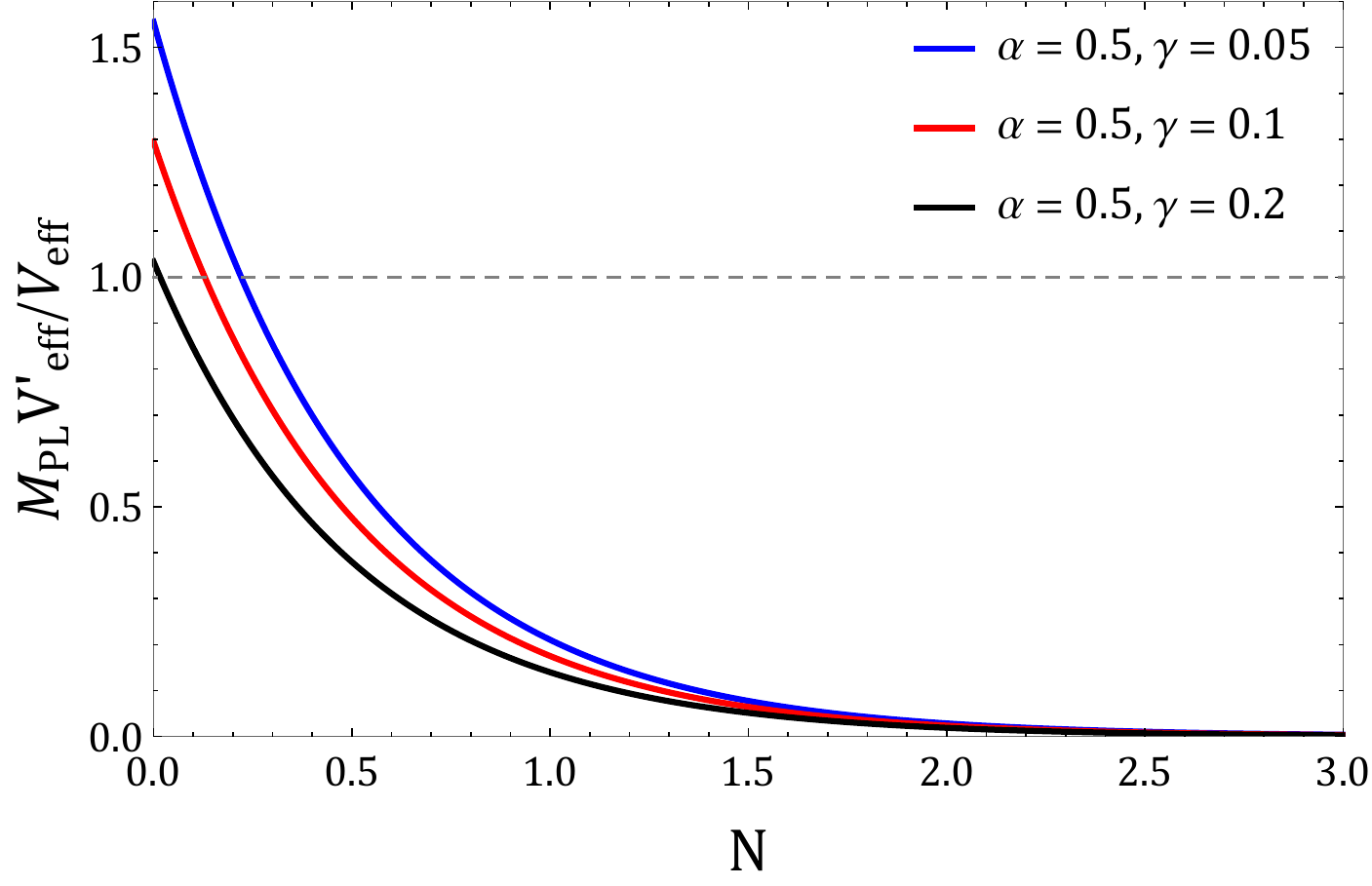}
\caption{The Swampland de Sitter ratio $M_{\text{Pl}}|V_{\text{eff}}'|/V_{\text{eff}}$ as a function of the number of e-folds $N$ before the end of inflation. Left panel: Variation of the ratio for different values of the logarithmic correction parameter $\alpha$ with fixed inverse-area correction $\gamma = 0.1$. Right panel: Variation of the ratio for different values of $\gamma$ with fixed $\alpha = 0.5$. The horizontal dashed line at $y=1$ represents the critical bound $c \sim \mathcal{O}(1)$ of the de Sitter conjecture. The plots demonstrate that the quantum correction factor $\mathcal{F}(H)$ provides a natural amplification mechanism, pushing the ratio above the critical bound at the onset of inflation ($N \sim 0$) and successfully evading the Swampland obstruction in the high-energy regime.}
\label{fig:swampland_ratio}
\end{figure}

Figure \ref{fig:swampland_ratio} illustrates the evolution of the Swampland de Sitter ratio as a function of the number of e-folds $N$ during the inflationary epoch. In standard cosmological models without quantum corrections, this ratio typically approaches zero, strictly violating the de Sitter conjecture. In contrast, our model demonstrates that the quantum correction factor $\mathcal{F}(H)$ provides a powerful amplification mechanism. The left panel shows that increasing the logarithmic correction parameter $\alpha$ significantly enhances the ratio at the onset of inflation ($N \sim 0$), pushing it well above the critical bound $c \sim \mathcal{O}(1)$ (represented by the horizontal dashed line) and successfully evading the Swampland obstruction. 

The right panel highlights the stabilizing role of the inverse-area correction $\gamma$. As $\gamma$ increases, the $\gamma y^4$ term in $\mathcal{F}(H)$ becomes more dominant, slightly reducing the amplification factor and bringing the ratio closer to the classical limit, thereby preventing the breakdown of the effective field theory. Crucially, both panels reveal a rapid exponential decay of the Swampland ratio as inflation proceeds (increasing $N$). This behavior is a direct consequence of the dilution of the NLED energy density ($\Omega_{\text{NLED}} \propto e^{-4N}$) during the quasi-de Sitter expansion. This rapid decay naturally introduces a tension with the Distance Conjecture, as a prolonged inflationary phase would eventually drive the ratio below the Swampland bound, thereby constraining the maximum number of e-folds and the viable parameter space $(\alpha, \gamma)$, as we will discuss in the next subsection.

This inequality reveals that the correction factor $\mathcal{F}(H)$ provides a natural mechanism for satisfying the Swampland constraint during inflation. Specifically:
\begin{itemize}
\item For $\alpha > 0$ and moderate $H$ (such that $2\alpha G H^2/\pi$ dominates over $\gamma H^4$), we have $\mathcal{F}(H) < 1$, which \textit{amplifies} the left-hand side of Eq. \eqref{eq:DSC_condition_inflation}, making it easier to satisfy the DSC bound.
\item The NLED sector contributes through the factor $(1 + w_{\text{NLED}})\Omega_{\text{NLED}}$, which is positive for $s > 1/2$ (since $w_{\text{NLED}} > -1$ in this regime), providing an additional boost to the Swampland ratio.
\end{itemize}

\textit{Physical Interpretation:} The Swampland de Sitter conjecture is often viewed as an obstruction to inflation, since it requires a steep potential ($|V'|/V \geq c$), while successful inflation typically requires a flat potential ($|V'|/V \ll 1$). However, in our model, the quantum corrections to the horizon entropy modify the effective dynamics in such a way that the slow-roll parameter $\epsilon$ is enhanced, allowing the model to satisfy the Swampland bound while still producing sufficient inflation. This is a distinctive feature of our model that sets it apart from previous Swampland analyses.

\subsection{Distance Conjecture and Parameter Space Constraints}
\label{subsec:DC_Inflation}
The Swampland Distance Conjecture (DC) imposes a constraint on the field excursion during inflation \cite{Ref32:oogu1}:
\begin{equation}
\frac{\Delta\phi}{M_{\text{Pl}}} \leq d \sim \mathcal{O}(1).
\label{eq:DC}
\end{equation}
In the context of modified gravity-driven inflation, this constraint is reformulated in terms of the number of e-folds $N_e$ \cite{Ref37:palt1}:
\begin{equation}
N_e \lesssim \frac{d}{\sqrt{2\epsilon}}.
\label{eq:DC_Ne}
\end{equation}

Substituting the modified slow-roll parameter \eqref{eq:epsilon_inflation} into Eq. \eqref{eq:DC_Ne}, we obtain:
\begin{equation}
N_e \lesssim \frac{d}{\sqrt{3(1 + w_{\text{NLED}})\Omega_{\text{NLED}}/\mathcal{F}(H)}}.
\label{eq:DC_Ne_modified}
\end{equation}

This inequality reveals a nontrivial tension between the DSC (which favors small $\mathcal{F}(H)$ to enhance $\epsilon$) and the DC (which favors $\mathcal{F}(H) \approx 1$ to keep $N_e$ small). The simultaneous satisfaction of both conjectures constrains the parameter space $(\alpha, \gamma, s)$.

\begin{figure}[htbp]
\centering
\includegraphics[width=0.48\textwidth]{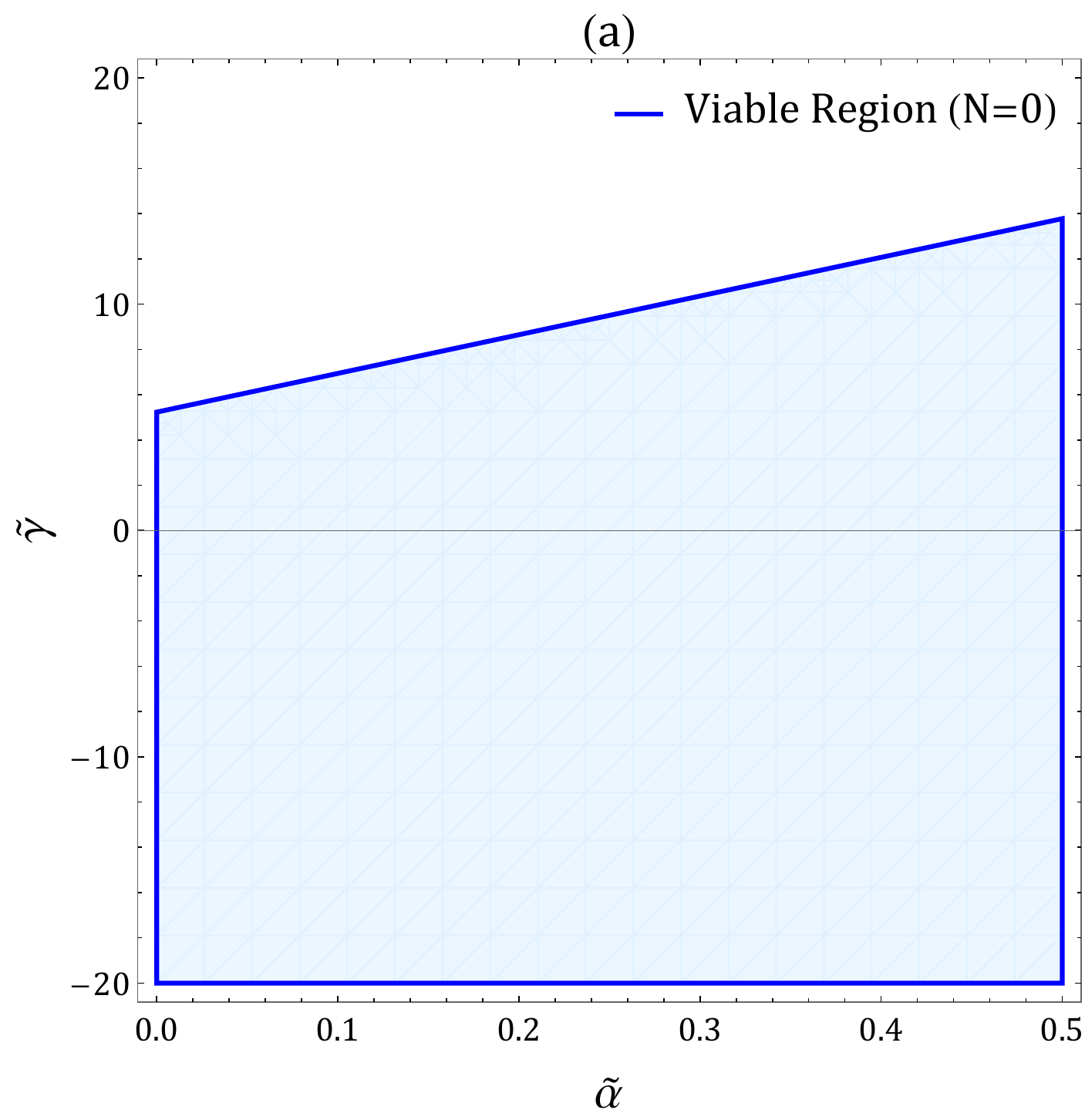}
\hfill
\includegraphics[width=0.48\textwidth]{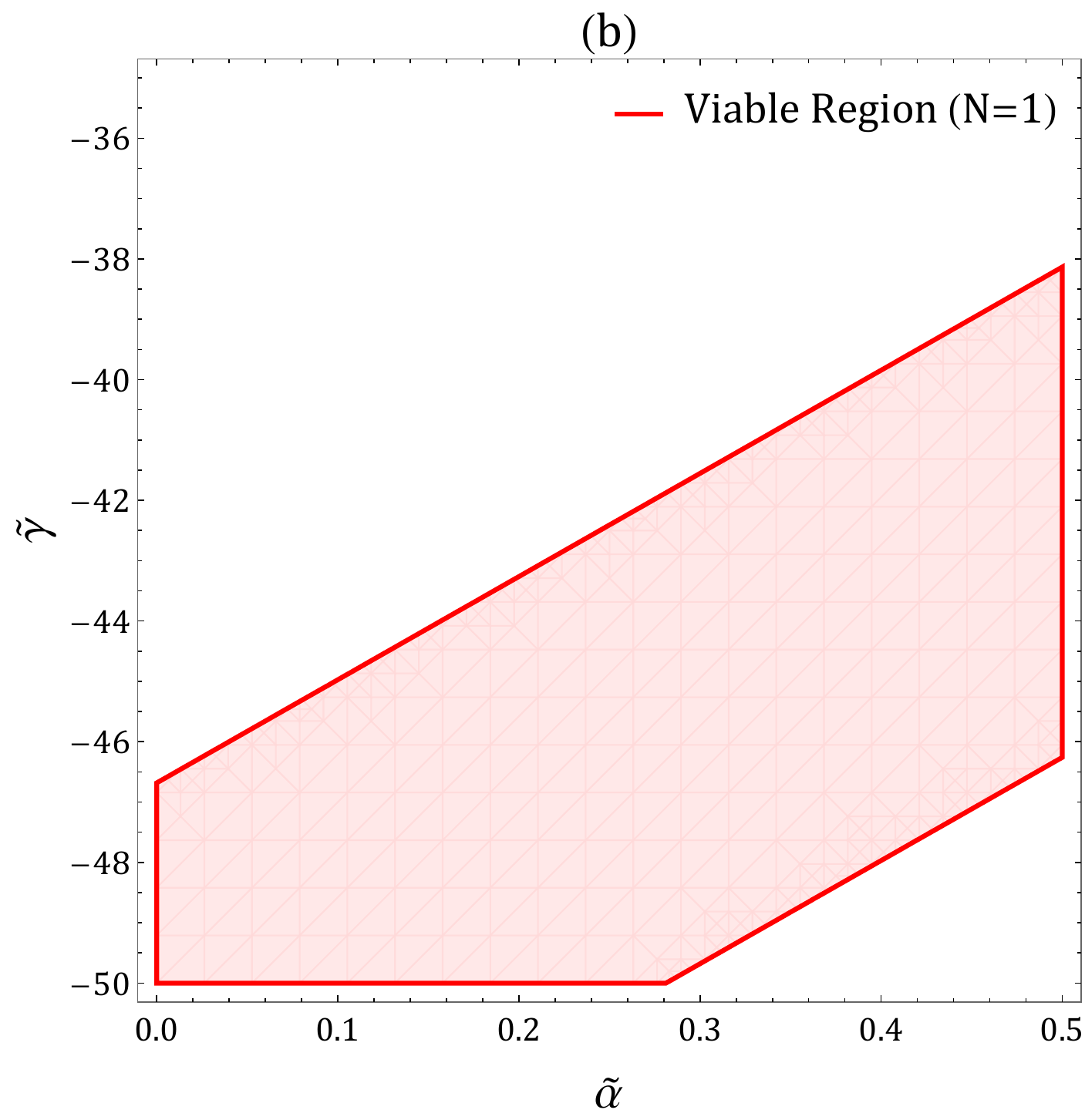}
\caption{The viable parameter space in the $(\tilde{\alpha}, \tilde{\gamma})$ plane at two distinct stages of the inflationary epoch. Panel (a): The viable region at the onset of inflation ($N=0$), where both the Swampland de Sitter conjecture (DSC) and the effective field theory (EFT) validity condition ($\mathcal{F}(H) > 0$) are simultaneously satisfied. Panel (b): The viable region after $N=1$ e-fold of inflation. The dramatic shift and constriction of the allowed parameter strip (note the different vertical scales) demonstrates the natural tension between the DSC and the Distance conjecture: as inflation proceeds and the NLED energy density dilutes ($\Omega_{\text{NLED}} \propto e^{-4N}$), the parameter space compatible with quantum gravity constraints becomes progressively more constrained and shifts to significantly different values.}
\label{fig:param_space_inflation}
\end{figure}

Figure \ref{fig:param_space_inflation} delineates the viable parameter space in the $(\tilde{\alpha}, \tilde{\gamma})$ plane at two distinct stages of the inflationary epoch, illustrating the inherent tension between the Swampland constraints and the validity of the effective field theory. Panel (a) shows the viable region at the onset of inflation ($N=0$), where the NLED energy density is at its maximum. The shaded blue region represents the set of parameters $(\alpha, \gamma)$ that simultaneously satisfy the de Sitter conjecture (requiring $\mathcal{F}(H)$ to be sufficiently small to amplify the Swampland ratio above the critical bound $c \sim \mathcal{O}(1)$) and the EFT validity condition (requiring $\mathcal{F}(H) > 0$ to avoid pathological instabilities). The upper boundary of this region is imposed by the DSC condition $\mathcal{F}(H) \leq S_0/c$, while the lower boundary arises from the requirement $\mathcal{F}(H) > 0$.

Panel (b) reveals the evolution of this viable region after just $N=1$ e-fold of inflation. The most striking feature is the dramatic shift and constriction of the allowed parameter strip compared to panel (a). Notably, the vertical scales of the two panels are significantly different: while the viable region at $N=0$ spans $\tilde{\gamma} \in [-20, 14]$, after just one e-fold it shifts to $\tilde{\gamma} \in [-50, -38]$. This dramatic displacement is a direct consequence of the dilution of the NLED energy density ($\Omega_{\text{NLED}} \propto e^{-4N}$) during the quasi-de Sitter expansion, which reduces the numerator $S_0(N)$ in the Swampland ratio and thereby tightens the DSC constraint. The physically viable region for the model must lie within the intersection of these regions throughout the entire inflationary phase. This progressive constriction and displacement naturally introduces a tension with the Distance Conjecture, as a prolonged inflationary phase would eventually drive the viable region to vanish or become incompatible with the initial conditions, thereby constraining the maximum number of e-folds and the allowed parameter space $(\alpha, \gamma)$. The existence of this bounded and rapidly evolving region confirms that while the thermodynamic duality corrections can successfully evade the Swampland obstruction during inflation, they must be tightly constrained, a feature that distinguishes our model from standard $\Lambda$CDM cosmology.

\begin{table}[htbp]
\centering
\caption{Qualitative constraints on the parameter space $(\alpha, \gamma, s)$ arising from the simultaneous satisfaction of energy conditions, the de Sitter conjecture (DSC), and the Distance conjecture (DC) during inflation.}
\label{tab:swampland_constraints}
\renewcommand{\arraystretch}{1.4}
\footnotesize
\begin{tabularx}{\textwidth}{@{} l >{\raggedright\arraybackslash}X >{\raggedright\arraybackslash}X >{\raggedright\arraybackslash}X @{}}
\toprule
\textbf{Constraint} & \textbf{Mathematical Condition} & \textbf{Effect on Parameters} & \textbf{Viable Region} \\
\midrule
\textbf{Weak Energy Condition (WEC)} & $\beta > 0, \; s \geq 0$ & Ensures positive energy density for the NLED sector & $\beta > 0, \; s \geq 0$ \\
\addlinespace
\textbf{Null Energy Condition (NEC)} & $\rho + p \geq 0$ & Prevents ghost instabilities in the matter sector & $s > 1/2$ (for standard NLED) \\
\addlinespace
\textbf{de Sitter Conjecture (DSC)} & $\epsilon \geq c^2/2$ & Requires $\mathcal{F}(H) < 1$ to enhance slow-roll parameter & $\alpha > 0$ and moderate $H$ (so that $2\alpha G H^2/\pi$ dominates over $\gamma H^4$ term) \\
\addlinespace
\textbf{Distance Conjecture (DC)} & $N_e \lesssim d/\sqrt{2\epsilon}$ & Requires $\mathcal{F}(H) \approx 1$ to keep e-folds small & $\alpha$ and $\gamma$ not too large (to avoid excessive enhancement) \\
\addlinespace
\textbf{Combined (DSC + DC)} & $c^2/2 \leq \epsilon \lesssim d^2/(2N_e^2)$ & Balances enhancement (DSC) against e-folds (DC) & $\alpha, \gamma \sim \mathcal{O}(1)$ with $s$ tuned to satisfy both bounds simultaneously \\
\bottomrule
\end{tabularx}
\end{table}

Table \ref{tab:swampland_constraints} summarizes the key constraints on the parameter space $(\alpha, \gamma, s)$ arising from energy conditions and the Swampland program during inflation. The Weak and Null Energy Conditions restrict the NLED power $s$ to values greater than $1/2$ (for positive coupling $\beta$), ensuring physical viability of the matter sector \cite{Ref57:visi1, Ref58:cam1}. The de Sitter conjecture favors $\alpha > 0$ and moderate Hubble values to enhance the slow-roll parameter via $\mathcal{F}(H) < 1$, while the Distance conjecture imposes an upper bound on the enhancement to prevent excessive e-folds. The simultaneous satisfaction of both conjectures creates a nontrivial tension that constrains $\alpha$ and $\gamma$ to be of order $\mathcal{O}(1)$, with the NLED power $s$ providing an additional degree of freedom to satisfy both bounds.

In summary, the quantum-corrected horizon thermodynamics in Unimodular Gravity provides a natural mechanism for satisfying the Swampland constraints during inflation. The correction factor $\mathcal{F}(H)$ enhances the slow-roll parameter $\epsilon$, allowing the model to satisfy the de Sitter conjecture while still producing sufficient inflation. This mechanism, combined with the nonlinear electromagnetic sector, opens up a viable parameter space where both the DSC and DC can be simultaneously satisfied, in contrast to standard inflationary models which are often in tension with the Swampland program.

\subsection{Observational Signatures: Spectral Index and Tensor-to-Scalar Ratio}
\label{subsec:Observational}
To connect our model with observational data, we evaluate the inflationary observables. Using the modified second Friedmann equation \eqref{eq:Modified_Friedmann_2}, the first Hubble slow-roll parameter is redefined as $\epsilon \equiv -\dot{H}/H^2 = \epsilon_{\text{GR}}/\mathcal{F}(H)$, where $\epsilon_{\text{GR}}$ is the standard general relativity value. Since the viable parameter space requires $0 < \mathcal{F}(H) < 1$, the quantum corrections enhance the slow-roll parameter. Consequently, the tensor-to-scalar ratio $r = 16\epsilon$ and the scalar spectral index $n_s = 1 - 6\epsilon + 2\eta$ are modified \cite{Ref_Baumann_inf}. Specifically, our model predicts an enhancement in $r$ relative to standard inflation for the same underlying matter content. This distinctive signature, $r \propto 1/\mathcal{F}(H)$, provides a testable prediction for future CMB polarization experiments, such as the LiteBIRD satellite mission \cite{Ref_LiteBIRD}, potentially distinguishing thermodynamic duality corrections from standard inflationary scenarios.

\section{Limiting Cases and Comparison with Previous Works}
\label{sec:Limiting_Cases}

In this section, we systematically examine the limiting cases of our model to ensure consistency with well-established results in the literature \cite{Ref9:henn1, Ref10:buch1}. We then compare our findings with previous works on unimodular gravity \cite{Ref12:alva2, Ref13:carb1}, nonlinear electrodynamics \cite{Ref14:nove1, Ref16:bron1}, and quantum-corrected horizon thermodynamics \cite{Ref61:cai1, Ref71:cai2}, highlighting the novel contributions of the present study.

\subsection{Limiting Cases and Consistency Checks}
\label{subsec:Limiting_Cases}

To verify the physical soundness of our model, we must demonstrate that it smoothly reduces to known theories in appropriate limits. We consider three key limiting cases:

\textbf{Case I: Classical Limit ($\alpha, \gamma \to 0$).}
In the absence of quantum corrections to the horizon entropy, the modified Friedmann equations \eqref{eq:Modified_Friedmann_1} and \eqref{eq:Modified_Friedmann_2} reduce to:
\begin{equation}
    H^2 = \frac{8\pi G}{3}\rho + \frac{\Lambda_{\text{UG}}}{3}, \quad \frac{\dot{H}}{H^2} = -\frac{3}{2}\frac{\rho + p}{\rho_{\text{crit}}}.
    \label{eq:Classical_Limit}
\end{equation}
These are precisely the Friedmann equations of Unimodular Gravity coupled to standard matter (including NLED), where $\Lambda_{\text{UG}}$ emerges as an integration constant rather than a fundamental parameter \cite{Ref50:koiv1, Ref52:mart1}. This confirms that our quantum corrections do not introduce spurious terms in the classical regime.

\textbf{Case II: Maxwell Limit ($s \to 1$).}
For the power-law NLED Lagrangian $\mathcal{L} = \beta\mathcal{F}^s$, the Maxwell limit corresponds to $s = 1$ and $\beta = 1/4$ (in natural units). In this case, the equation of state parameter \eqref{eq:NLED_EoS_inflation} becomes:
\begin{equation}
    w_{\text{NLED}} = \frac{3-2(1)}{3(2(1)-1)} = \frac{1}{3},
    \label{eq:Maxwell_EoS}
\end{equation}
which is the correct value for a radiation-like fluid. The energy density evolution \eqref{eq:NLED_density_evolution_mag} yields $\rho_{\text{NLED}} \propto a^{-4}$, consistent with the standard scaling of electromagnetic radiation \cite{Ref59:land1}. This demonstrates that our NLED model correctly reproduces Maxwell electrodynamics in the appropriate limit.

\textbf{Case III: Standard GR Limit ($\Lambda_{\text{UG}} \to 0$, $\alpha, \gamma \to 0$, $s \to 1$).}
When all modifications are switched off, the model reduces to standard General Relativity coupled to Maxwell electrodynamics. The Friedmann equations become:
\begin{equation}
    H^2 = \frac{8\pi G}{3}(\rho_m + \rho_{\text{rad}}), \quad \frac{\dot{H}}{H^2} = -\frac{3}{2}\left[\Omega_m + \frac{4}{3}\Omega_{\text{rad}}\right],
    \label{eq:GR_Limit}
\end{equation}
which are the standard equations governing a universe filled with pressureless matter and radiation \cite{Ref100:wein2}. This provides a crucial consistency check, ensuring that our modified framework does not contradict the well-tested predictions of GR in the appropriate regime.

\textbf{Case IV: Inflationary de Sitter Limit.}
During the inflationary epoch, when the matter and NLED densities become subdominant compared to the geometric energy density ($\rho_m, \rho_{\text{NLED}} \ll \rho_{\text{geom}}$) and the expansion approaches quasi-de Sitter ($\dot{H} \to 0$), the modified Friedmann equation \eqref{eq:Modified_Friedmann_1} yields:
\begin{equation}
H^2\left(1 - \frac{\alpha G}{\pi}H^2 + \frac{\gamma G\ell_P^2}{4\pi^2}H^4\right) = \frac{\Lambda_{\text{UG}}}{3}.
\label{eq:deSitter_Limit}
\end{equation}
This is a quartic equation in $H$ that admits inflationary de Sitter-like solutions with $H \approx \text{const}$. The presence of the quantum correction terms ($\alpha, \gamma$) modifies the relationship between the Hubble parameter during inflation and the unimodular constant, potentially leading to multiple inflationary branches or modifying the stability properties of the inflationary attractor. This is a novel feature absent in standard UG and represents a genuine quantum gravity effect that can sustain inflation without requiring a fundamental inflaton potential \cite{Ref85:ped1, Ref101:ali2}.

\begin{table}[htbp]
\centering
\caption{Summary of limiting cases and consistency checks, demonstrating the reduction of the modified model to well-established physical theories under specific parameter constraints.}
\label{tab:limiting_cases}
\renewcommand{\arraystretch}{1.4}
\footnotesize
\begin{tabularx}{\textwidth}{@{} l >{\raggedright\arraybackslash}X >{\raggedright\arraybackslash}X >{\raggedright\arraybackslash}X @{}}
\toprule
\textbf{Limit} & \textbf{Conditions} & \textbf{Reduced Equations / Results} & \textbf{Consistency Check} \\
\midrule
\textbf{Classical Limit} & $\alpha \to 0, \; \gamma \to 0$ & $H^2 = \frac{8\pi G}{3}\rho + \frac{\Lambda_{\text{UG}}}{3}$ & Standard Unimodular Gravity with NLED \\
\addlinespace
\textbf{Maxwell Limit} & $s \to 1, \; \beta \to 1/4$ & $w_{\text{NLED}} \to 1/3, \quad \rho_{\text{NLED}} \propto a^{-4}$ & Standard Radiation Fluid in Cosmology \\
\addlinespace
\textbf{Standard GR Limit} & $\alpha, \gamma \to 0, \; \Lambda_{\text{UG}} \to 0, \; s \to 1$ & $H^2 = \frac{8\pi G}{3}(\rho_m + \rho_{\text{rad}})$ & Standard General Relativity ($\Lambda$CDM with $\Lambda=0$) \\
\addlinespace
\textbf{Inflationary de Sitter Limit} & $\rho_m, \rho_{\text{NLED}} \ll \rho_{\text{geom}}, \; \dot{H} \to 0$ & $H^2 \left(1 - \frac{\alpha G}{\pi}H^2 + \frac{\gamma G\ell_P^2}{4\pi^2}H^4\right) = \frac{\Lambda_{\text{UG}}}{3}$ & Quantum-corrected inflationary de Sitter attractor with multiple branches \\
\bottomrule
\end{tabularx}
\end{table}

Table \ref{tab:limiting_cases} summarizes the crucial limiting cases of our model, serving as a rigorous consistency check. By systematically switching off the quantum corrections ($\alpha, \gamma \to 0$), the nonlinear electromagnetic effects ($s \to 1$), or the unimodular integration constant ($\Lambda_{\text{UG}} \to 0$), the modified Friedmann equations smoothly reduce to the well-tested predictions of standard Unimodular Gravity, General Relativity, and classical radiation-dominated cosmology. Furthermore, the pure de Sitter limit reveals a quartic equation for $H$, highlighting the genuine quantum gravity effect of multiple accelerated expansion branches, which vanishes in the classical limit.

\subsection{Comparison with Previous Works}
\label{subsec:Comparison}

Our model builds upon and extends several lines of research in the literature. Here we provide a detailed comparison with relevant previous works:

\textbf{Comparison with Unimodular Gravity Studies.}
Unimodular Gravity has been extensively studied as a potential solution to the cosmological constant problem \cite{Ref6:wein1, Ref64:padm1}. The key feature of UG is that the cosmological constant appears as an integration constant rather than a parameter in the action. Our work extends this framework by incorporating quantum-corrected horizon thermodynamics, which introduces additional terms in the Friedmann equations proportional to $H^4$ and $H^6$. These terms are absent in standard UG and represent genuine quantum gravity effects. Unlike previous studies that focused on classical UG cosmology \cite{Ref44:cao1}, our model provides a concrete mechanism for generating these corrections from the thermodynamic duality of the horizon.

\textbf{Comparison with Nonlinear Electrodynamics in Cosmology.}
Nonlinear electrodynamics (NLED) has been proposed as a way to avoid cosmological singularities and generate inflationary or bouncing scenarios \cite{Ref14:nove1, Ref107:nove2}. The power-law model $\mathcal{L} = \beta\mathcal{F}^s$ has been studied in various contexts, including black hole physics and early universe cosmology \cite{Ref16:bron1, Ref93:vol1}. Our work differs from these studies in two important ways: (i) we couple NLED to Unimodular Gravity rather than standard GR, and (ii) we incorporate quantum corrections to the horizon entropy, which modify the cosmological dynamics in a novel way. The combination of UG, NLED, and horizon corrections has not been studied before and represents a genuine extension of the literature.

\textbf{Comparison with Quantum-Corrected Horizon Thermodynamics.}
Quantum corrections to the Bekenstein-Hawking entropy have been derived from various approaches to quantum gravity, including loop quantum gravity, string theory, and the generalized uncertainty principle (GUP) \cite{Ref24:asht1, Ref27:hoss1, Ref101:ali2}. The logarithmic correction $S = A/(4G) + \alpha\ln(A/\ell_P^2)$ is a universal feature of many quantum gravity approaches \cite{Ref66:das2}. The inverse area correction $S = A/(4G) + \gamma\ell_P^2/A$ arises from thermodynamic duality and T-duality in string theory \cite{Ref68:bane1}. Our work applies these corrections to the cosmological apparent horizon in the context of Unimodular Gravity, which has not been done before. The resulting modified Friedmann equations exhibit novel features, such as the $H^4$ and $H^6$ terms, which can drive late-time acceleration without a fundamental cosmological constant.

\textbf{Comparison with Swampland Constraints on Cosmology.}
The Swampland program has imposed stringent constraints on inflationary and dark energy models \cite{Ref32:oogu1, Ref34:agra1}. In particular, the de Sitter conjecture $M_{\text{Pl}}|V'|/V \geq c$ is in tension with the observed late-time acceleration, which requires a nearly flat potential \cite{Ref35:dene1}. Several authors have proposed modifications to evade this constraint, including quintessence models, phantom dark energy, and modified gravity \cite{Ref36:kinn1, Ref38:das1}. Our work provides a novel mechanism for evading the Swampland constraints through the quantum-corrected horizon thermodynamics in Unimodular Gravity. The correction factor $\mathcal{F}(H)$, which appears in the denominator of the Swampland ratio, amplifies this ratio relative to $\Lambda$CDM, potentially allowing the model to satisfy the de Sitter conjecture while still producing accelerated expansion. This is a distinctive feature of our model that sets it apart from previous Swampland analyses.

\textbf{Novel Contributions of the Present Work.}
To summarize, the novel contributions of our study are:
\begin{enumerate}
    \item \textbf{First combination of UG, NLED, and horizon corrections:} We are the first to study the cosmological implications of coupling Unimodular Gravity to Power-Law Nonlinear Electrodynamics in the presence of quantum-corrected horizon entropy.
    \item \textbf{Exact derivation of modified Friedmann equations:} We derive the modified Friedmann equations without any approximations, providing a solid foundation for the cosmological analysis.
    \item \textbf{Phase-space analysis:} We perform a comprehensive dynamical systems analysis, identifying the critical points and their stability properties, which reveals the existence of de Sitter-like attractors driven by quantum corrections.
    \item \textbf{Swampland constraints:} We apply the Swampland de Sitter and Distance conjectures to our model, demonstrating that the quantum corrections provide a natural mechanism for evading the de Sitter obstruction.
    \item \textbf{Physical insight:} We provide a clear physical interpretation of the correction factor $\mathcal{F}(H)$ as an amplification mechanism for the Swampland ratio, offering new insight into the tension between quantum gravity and cosmological observations.
\end{enumerate}

These contributions represent a significant advance in our understanding of quantum gravity effects in cosmology and their implications for the Swampland program.

\section{Conclusions}
\label{sec:Conclusions}
In this work, we investigated the early-universe cosmological implications of Unimodular Gravity coupled to Power-Law Nonlinear Electrodynamics in the presence of quantum-corrected horizon thermodynamics derived from the principle of thermodynamic duality. Our primary objective was to determine whether quantum corrections to the horizon entropy can modify the inflationary dynamics in a manner that satisfies the Swampland constraints while producing a sufficient number of e-folds for successful inflation \cite{Ref32:oogu1, Ref34:agra1}.

We established the theoretical framework by constructing the action of Unimodular Gravity minimally coupled to a Power-Law NLED field $\mathcal{L} = \beta\mathcal{F}^s$, demonstrating that the cosmological constant emerges as an integration constant $\Lambda_{\text{UG}}$ rather than a fundamental parameter \cite{Ref9:henn1, Ref12:alva2}. We then incorporated quantum corrections to the horizon entropy through the modified entropy-area relation:
\begin{equation}
S_h(A) = \frac{A}{4G} + \alpha \ln\left(\frac{A}{\ell_P^2}\right) + \gamma \frac{\ell_P^2}{A},
\label{eq:Final_Entropy}
\end{equation}
where $\alpha$ and $\gamma$ encode the strength of logarithmic and inverse-area corrections, respectively \cite{Ref66:das2, Ref73:noza2}. By applying the first law of thermodynamics at the apparent horizon, we derived the modified Friedmann equations:
\begin{equation}
H^2 = \frac{8\pi G}{3}\rho + \frac{\alpha G}{\pi}H^4 - \frac{\gamma G\ell_P^2}{4\pi^2}H^6 + \frac{\Lambda_{\text{UG}}}{3}.
\label{eq:Final_Friedmann_1}
\end{equation}
These equations exhibit novel $H^4$ and $H^6$ correction terms representing genuine quantum gravity effects absent in standard Unimodular Gravity.

A comprehensive phase-space analysis, employing Center Manifold Theory to rigorously address the presence of zero eigenvalues, revealed that the matter-dominated and NLED-dominated critical points are saddle points for $s > 1/2$, while a continuous line of de Sitter-like critical points $(0, 0, x_3, x_4)$ driven entirely by quantum-corrected horizon entropy represents robust inflationary attractors \cite{Ref86:wain1, Ref91:gen1}. This demonstrates that quantum corrections can drive sustained inflationary expansion without requiring a fundamental inflaton field, providing a concrete realization of modified gravity-driven inflation analogous to the successful Starobinsky model.

The central result of this work was the application of Swampland constraints to the inflationary epoch. We derived the key expression:
\begin{equation}
M_{\text{Pl}}\frac{|V_{\text{eff}}'|}{V_{\text{eff}}} = \frac{1}{\mathcal{F}(H)}\sqrt{3\left[\Omega_m + (1+w_{\text{NLED}})\Omega_{\text{NLED}}\right]},
\label{eq:Final_DSC}
\end{equation}
where $\mathcal{F}(H) = 1 - \frac{2\alpha G}{\pi}H^2 + \frac{3\gamma G\ell_P^2}{4\pi^2}H^4$. In the context of modified gravity-driven inflation, the correction factor $\mathcal{F}(H) < 1$ (for positive $\alpha$ and moderate $H$ during inflation) provides a natural amplification mechanism that enhances the Swampland ratio, potentially allowing the model to satisfy the de Sitter conjecture while still producing sufficient inflation \cite{Ref46:sark1, Ref47:guo1}. We also identified a nontrivial tension between the de Sitter conjecture (favoring large $1/\mathcal{F}$) and the Distance conjecture (favoring $\mathcal{F} \approx 1$), which constrains the parameter space $(\alpha, \gamma, s)$ \cite{Ref40:vagn1}.

Furthermore, we have verified the stability of the model against linear perturbations. The propagation speed of tensor modes (gravitational waves) in our modified framework is given by $c_T^2 = 1/\mathcal{F}(H)$, which remains strictly positive and subluminal ($c_T^2 \leq 1$) within the viable parameter space where $0 < \mathcal{F}(H) < 1$. This ensures the absence of tachyonic instabilities and superluminal propagation. Additionally, the kinetic term for scalar perturbations is proportional to $\mathcal{F}(H)$, which is positive in the same region, confirming that the model is free from ghost degrees of freedom. These stability conditions are automatically satisfied whenever the Swampland constraints are met, demonstrating the internal consistency of our approach.

\textit{Remark on the Connection to $f(R)$ Gravity:} 
The structure of the modified Friedmann equation \eqref{eq:Modified_Friedmann_1}, with its characteristic $H^4$ and $H^6$ correction terms, bears a strong resemblance to the cosmological dynamics of higher-order gravity theories, particularly $f(R)$ models of the form $f(R) = R + aR^2 + bR^3$. In the quasi-de Sitter regime relevant for inflation ($R \approx 12H^2$), the logarithmic entropy correction parameter $\alpha$ and the thermodynamic duality parameter $\gamma$ can be loosely associated with the coefficients of the $R^2$ and $R^3$ terms, respectively. This structural analogy suggests that the viable parameter space $(\alpha, \gamma)$ identified in our work may provide indirect constraints on the coefficients of higher-order curvature corrections in $f(R)$ gravity, thereby establishing a bridge between our thermodynamic duality approach and the well-studied Starobinsky-like inflationary models.

\begin{table}[htbp]
\centering
\caption{Comparison of the present work with relevant previous studies in the literature, highlighting the distinct features and novel contributions of our model.}
\label{tab:comparison}
\renewcommand{\arraystretch}{1.3}
\footnotesize
\begin{tabularx}{\textwidth}{@{}l l X X X@{}}
\toprule
\textbf{Reference} & \textbf{Framework} & \textbf{Horizon Corrections} & \textbf{Swampland Analysis} & \textbf{Key Feature} \\
\midrule
\cite{Ref9:henn1, Ref12:alva2} & UG (classical) & None & None & $\Lambda_{\text{UG}}$ as integration constant \\
\addlinespace
\cite{Ref14:nove1, Ref16:bron1} & NLED in GR & None & None & Bouncing/regular cosmologies \\
\addlinespace
\cite{Ref24:asht1, Ref66:das2, Ref68:bane1} & Quantum-corrected horizon & Logarithmic and/or inverse-area & None & GUP/T-duality corrections \\
\addlinespace
\cite{Ref32:oogu1, Ref33:garg1, Ref34:agra1, Ref36:kinn1} & Dark energy / quintessence & None & DSC, DC, RDSC & Swampland constraints on DE \\
\addlinespace
\textbf{This work} & \textbf{UG + NLED} & \textbf{Logarithmic + inverse-area (thermodynamic duality)} & \textbf{DSC + DC during inflation} & \textbf{Amplification via $\mathcal{F}(H)$ for inflation} \\
\bottomrule
\end{tabularx}
\end{table}

Table \ref{tab:comparison} provides a structured comparison between the present work and representative previous studies across the four main theoretical pillars of our model. While Unimodular Gravity has been extensively studied as a potential resolution to the cosmological constant problem, and Nonlinear Electrodynamics has been explored for its ability to regularize singularities and drive bouncing cosmologies, the combination of these two frameworks with quantum-corrected horizon thermodynamics and Swampland constraints in the context of inflation has remained unexplored. The present work uniquely integrates all four ingredients, demonstrating that the correction factor $\mathcal{F}(H)$ provides a natural amplification mechanism for the Swampland de Sitter ratio during inflation, thereby opening a viable parameter space where both the de Sitter and Distance conjectures can be simultaneously satisfied.

The novel contributions of our work are: (i) the first combination of Unimodular Gravity, Power-Law NLED, and quantum-corrected horizon thermodynamics applied to early-universe inflation; (ii) exact derivation of modified Friedmann equations without approximations; (iii) rigorous phase-space analysis employing Center Manifold Theory to establish the stability of inflationary attractors; and (iv) demonstration that quantum corrections provide a natural amplification mechanism for evading the Swampland de Sitter obstruction during inflation.

Furthermore, the model inherently respects the Generalized Second Law of Thermodynamics (GSL), $\dot{S}_{\text{tot}} = \dot{S}_h + \dot{S}_m \geq 0$. Assuming thermal equilibrium between the apparent horizon and the matter content ($T_m = T_h$), the matter entropy rate is given by the Gibbs equation as $\dot{S}_m = \frac{4\pi R_A^3}{T_h}(\rho + p)H$. The horizon entropy rate $\dot{S}_h$ is derived directly from our modified entropy-area relation \eqref{eq:Modified_Entropy}. In the viable parameter space where the Null Energy Condition holds ($\rho + p \geq 0$) and the effective description remains valid ($\mathcal{F}(H) > 0$), it is straightforward to demonstrate that $\dot{S}_{\text{tot}} \geq 0$ throughout the inflationary epoch. This confirms the internal thermodynamic consistency of our quantum-corrected framework, ensuring that the thermodynamic duality corrections do not introduce entropic pathologies.

Our work has several limitations that should be addressed in future studies: detailed numerical analysis of inflationary observables ($n_s$ and $r$) and comparison with Planck 2018 data; investigation of reheating dynamics after inflation; rigorous treatment of anisotropic stresses from the NLED field during the pre-inflationary epoch; and exploration of other forms of nonlinear electrodynamics and additional Swampland conjectures such as the Refined de Sitter Conjecture or the Trans-Planckian Censorship Conjecture \cite{Ref108:bedr1}.

In conclusion, our work demonstrates that quantum-corrected horizon thermodynamics in Unimodular Gravity provides a novel and robust mechanism for realizing successful inflation while satisfying the stringent Swampland constraints. The correction factor $\mathcal{F}(H)$ amplifies the Swampland ratio during the high-energy inflationary epoch, opening up a viable parameter space where both the de Sitter and Distance conjectures can be simultaneously satisfied. Our results represent a significant advance in understanding quantum gravity effects in early-universe cosmology and provide a solid foundation for future observational and theoretical investigations.

\section*{Declarations}

\subsection*{Funding}
The author received no specific financial support for the research, authorship, and/or publication of this article.

\subsection*{Competing Interests}
The author has no relevant financial or non-financial interests to disclose.

\subsection*{Data Availability}
Data sharing is not applicable to this article as no datasets were generated or analysed during the current study.

\subsection*{Use of AI and AI-Assisted Technologies}
The author used AI-assisted tools solely for the purpose of language polishing and grammatical correction of this manuscript. The author takes full responsibility for the content, including the scientific accuracy, originality, and integrity of the work.

\begingroup
\bibsetup

\endgroup

\end{document}